\documentclass[prd,superscriptaddress,twocolumn,amssymb,amsmath,amsfonts,aps,nofootinbib]{revtex4-1}
\newcommand{\gx}{\textsc{GlueX}}
\usepackage{lineno}
\usepackage[pdftex]{graphicx}
\usepackage[colorlinks=true,allcolors=blue,bookmarks=true]{hyperref}
\usepackage{color}
\definecolor{orange}{RGB}{153,86,0}

\begin{document}
%
%
\title{\boldmath{An initial study of mesons and baryons containing strange quarks with \gx} \\ 
{\small  (A proposal to the 40$^\mathrm{th}$ Jefferson Lab Program Advisory Committee)}}

\date{May 2, 2013}
\author{A.~AlekSejevs}
\author{S.~Barkanova}
\affiliation{Acadia University, Wolfville, Nova Scotia, B4P 2R6, Canada} 
\author{M.~Dugger}
\author{B.~Ritchie}
\author{I.~Senderovich}
\affiliation{Arizona State University, Tempe, Arizona 85287, USA}
\author{E.~Anassontzis}
\author{P.~Ioannou}
\author{C.~Kourkoumeli}
\author{G.~Voulgaris}
\affiliation{University of Athens, GR-10680 Athens, Greece}
\author{N.~Jarvis}
\author{W.~Levine}
\author{P.~Mattione}
\author{W.~McGinley}
\author{C.~A.~Meyer}\thanks{Spokesperson}
\author{R.~Schumacher}
\author{M.~Staib}
\affiliation{Carnegie Mellon University, Pittsburgh, Pennsylvania 15213, USA}
\author{P.~Collins}
\author{F.~Klein}
\author{D.~Sober}
\affiliation{Catholic University of America, Washington, D.C.\ 20064, USA}
\author{D.~Doughty}
\affiliation{Christopher Newport University, Newport News, Virginia 23606, USA}
\author{A.~Barnes}
\author{R.~Jones}
\author{J.~McIntyre}
\author{F.~Mokaya}
\author{B.~Pratt}
\affiliation{University of Connecticut, Storrs, Connecticut 06269, USA}
\author{W.~Boeglin}
\author{L.~Guo}
\author{P.~Khetarpal}
\author{E.~Pooser}
\author{J.~Reinhold}
\affiliation{Florida International University, Miami, Florida 33199, USA}
\author{H.~Al Ghoul}
\author{S.~Capstick}
\author{V.~Crede}
\author{P.~Eugenio}
\author{A.~Ostrovidov}
\author{N.~Sparks}
\author{A.~Tsaris}
\affiliation{Florida State University, Tallahassee, Florida 32306, USA}
\author{D.~Ireland}
\author{K.~Livingston}
\affiliation{University of Glasgow, Glasgow G12 8QQ, United Kingdom}
\author{D.~Bennett}
\author{J.~Bennett}
\author{J.~Frye}
\author{M.~Lara}
\author{J.~Leckey}
\author{R.~Mitchell}
\author{K.~Moriya}
\author{M.~R.~Shepherd}\thanks{Deputy Spokesperson}
\author{A.~Szczepaniak}
\affiliation{Indiana University, Bloomington, Indiana 47405, USA}
\author{R.~Miskimen}
\author{A.~Mushkarenkov}
\affiliation{University of Massachusetts, Amherst, Massachusetts 01003, USA}
\author{B.~Guegan}
\author{J.~Hardin}
\author{J.~Stevens}
\author{M.~Williams}
\affiliation{Massachusetts Institute of Technology, Cambridge, Massachusetts 02139, USA}
\author{A.~Ponosov}
\author{S.~Somov}
\affiliation{MEPHI, Moscow, Russia}
\author{C.~Salgado}
\affiliation{Norfolk State University, Norfolk, Virginia 23504, USA}
\author{P.~Ambrozewicz}
\author{A.~Gasparian}
\author{R.~Pedroni}
\affiliation{North Carolina A\&T State University, Greensboro, North Carolina 27411, USA}
\author{T.~Black}
\author{L.~Gan}
\affiliation{University of North Carolina, Wilmington, North Carolina 28403, USA}
\author{S.~Dobbs}
\author{K.~K.~Seth}
\author{A.~Tomaradze}
\affiliation{Northwestern University, Evanston, Illinois, 60208, USA}
\author{J.~Dudek}
\affiliation{Old Dominion University, Norfolk, Virginia 23529, USA}
\affiliation{Thomas Jefferson National Accelerator Facility, Newport News, Virginia 23606, USA}
\author{F.~Close}
\affiliation{University of Oxford, Oxford OX1 3NP, United Kingdom}
\author{E.~Swanson}
\affiliation{University of Pittsburgh, Pittsburgh, Pennsylvania 15260, USA }
\author{S.~Denisov}
\affiliation{Institute for High Energy Physics, Protvino, Russia}
\author{G.~Huber}
\author{D.~Kolybaba}
\author{S.~Krueger}
\author{G.~Lolos}
\author{Z.~Papandreou}
\author{A.~Semenov}
\author{I.~Semenova}
\author{M.~Tahani}
\affiliation{University of Regina, Regina, SK S4S 0A2, Canada}
\author{W.~Brooks}
\author{H.~Hakobyan}
\author{S.~Kuleshov}
\author{O.~Soto}
\author{A.~Toro}
\author{I.~Vega}
\author{R.~White}
\affiliation{Universidad T\'ecnica Federico Santa Mar\'ia, Casilla 110-V Valpara\'iso, Chile}
\author{F.~Barbosa}
\author{E.~Chudakov}\thanks{Hall D Leader}
\author{H.~Egiyan}
\author{M.~Ito}
\author{D.~Lawrence}
\author{M.~McCaughan}
\author{M.~Pennington}
\author{L.~Pentchev}
\author{Y.~Qiang}
\author{E.~S.~Smith}
\author{A.~Somov}
\author{S.~Taylor}
\author{T.~Whitlatch}
\author{E.~Wolin}
\author{B.~Zihlmann}
\affiliation{Thomas Jefferson National Accelerator Facility, Newport News, Virginia 23606, USA}

\collaboration{The \gx~Collaboration}

\begin{abstract}

The primary motivation of the \gx~experiment is to search for and ultimately study the pattern of
gluonic excitations in the meson spectrum produced in $\gamma p$ collisions. Recent lattice QCD 
calculations predict a rich spectrum of hybrid mesons that have both exotic and non-exotic $J^{PC}$, 
corresponding to $q\bar{q}$ states ($q=u,$ $d,$ or $s$) coupled with a gluonic field.  A thorough 
study of the hybrid spectrum, including the identification of the isovector triplet, with charges 0 and 
$\pm1$, and both isoscalar members, $|s\bar{s}\rangle$ and $|u\bar{u}\rangle + |d\bar{d}\rangle$, 
for each predicted hybrid combination of $J^{PC}$, may only be achieved by conducting a systematic 
amplitude analysis of many different hadronic final states.  Detailed studies of the performance of 
the \gx~detector have indicated that identification of particular final states with
kaons is possible using the baseline detector configuration.  The efficiency of kaon detection
coupled with the relatively lower production cross section for particles containing hidden strangeness
will require a high intensity run in order for analyses of such states to be feasible.
We propose to collect a total of $200$ days of physics analysis data at an average intensity of 
$5\times 10^7$ tagged photons on target per second.  This data sample will provide 
an order of magnitude statistical improvement over the initial \gx~running, which will allow us to begin 
a program of studying mesons and baryons containing strange quarks.  In addition, the increased
intensity will permit us to study reactions that may have been statistically limited in the initial phases of GlueX. 
Overall, this will lead to a significant increase in the potential for \gx~to make key experimental advances in 
our knowledge of hybrid mesons and excited $\Xi$ baryons.

\end{abstract}

\maketitle

\section{Introduction and background}

A long-standing goal of hadron physics has been to understand how the quark 
and gluonic degrees of freedom that are present in the fundamental QCD Lagrangian 
manifest themselves in the spectrum of hadrons.   Of particular interest is how the 
gluon-gluon interactions might give rise to physical states with gluonic excitations.  
One class of such states is the hybrid meson, 
which can be naively thought of as a quark anti-quark pair coupled to a valence 
gluon ($q\bar{q}g$).  Recent lattice QCD calculations~\cite{Dudek:2011bn} predict a rich spectrum 
of hybrid mesons.  A subset of these hybrids has an unmistakable experimental signature:  
angular momentum ($J$), parity ($P$), and charge conjugation ($C$) that cannot be 
created from just a quark-antiquark pair.  Such states are called exotic hybrid mesons.  
The primary goal of the \gx~experiment in Hall~D is to search for and study these mesons.

A detailed overview of the motivation for the \gx~experiment as well as the design 
of the detector and beamline can be found in the initial proposal to the Jefferson Lab 
Program Advisory Committee (PAC) 30~\cite{pac30}, a subsequent PAC 36 update~\cite{pac36} 
and a conditionally approved proposal to PAC 39~\cite{pac39}.
While the currently-approved 120 days of beam time with the baseline detector configuration 
will allow \gx~an unprecedented opportunity to search for exotic hybrid mesons, the statistics 
that will be collected during this period will be inadequate for studying mesons or baryons 
containing strange quarks. These issues were addressed in our proposal to PAC 39~\cite{pac39},
where we proposed a complete package that would allow us to fully explore the strangeness
sector using a combination of new particle-identification capability and the full implementation
of our level-three (software) trigger to increase the data-rate capabilities of the experiment. This full 
functionality will ultimately be needed for the \gx~experiment to complete its primary goal of solidifying 
our experimental understanding of hybrids by identifying {\em patterns} of hybrid mesons, both 
isoscalar and isovector, exotic and non-exotic, that are embedded in the spectrum of conventional 
mesons. However, there are select final states containing strange particles that
can be studied with the baseline \gx{} equipment provided that the statistical precision
of the data set is sufficient.  This proposal focuses on those parts of the \gx{} program that can 
be addressed with the baseline hardware, but will be statistically limited in the 
currently-approved \gx{} running time.  The motivation and experimental context for these
studies is largely the same as presented in our PAC 39 proposal; we repeat it here for
completeness.

\subsection{Theoretical context}
\label{sec:theory}

Our understanding of how gluonic excitations manifest themselves within QCD 
is maturing thanks to recent results from lattice QCD. This numerical approach 
to QCD considers the theory on a finite, discrete grid of points in a manner that 
would become exact if the lattice spacing were taken to zero and the spatial extent 
of the calculation, {\it i.e.,} the ``box size," 
was made infinitely large. In practice, rather fine spacings and large boxes are used so that 
the systematic effect of this approximation should be small. The main limitation of
these calculations at present is the poor scaling of the numerical algorithms 
with decreasing quark mass - in practice most contemporary calculations use a 
range of artificially heavy light quarks and attempt to observe a trend as the light 
quark mass is reduced toward the physical value. Trial calculations at the physical 
quark mass have begun and regular usage is anticipated within a few years.

The spectrum of eigenstates of QCD can be extracted from correlation functions 
of the type $\langle 0 | \mathcal{O}_f(t) \mathcal{O}_i^\dag(0) | 0 \rangle$, where 
the $\mathcal{O}^\dag$ are composite QCD operators capable of interpolating a 
meson or baryon state from the vacuum. The time-evolution of the Euclidean 
correlator indicates the mass spectrum ($e^{-m_\mathfrak{n} t}$) and information 
about quark-gluon substructure can be inferred from matrix-elements 
$\langle \mathfrak{n} | \mathcal{O}^\dag |0 \rangle$. In a series of recent 
papers~\cite{Dudek:2009qf,Dudek:2010wm,Dudek:2011tt,Edwards:2011jj}, 
the Hadron Spectrum Collaboration has explored the spectrum of mesons and 
baryons using a large basis of composite QCD interpolating fields, extracting a 
spectrum of states of determined $J^{P(C)}$, including states of high internal excitation.

\begin{figure*}
\begin{center}
\includegraphics[width=\linewidth]{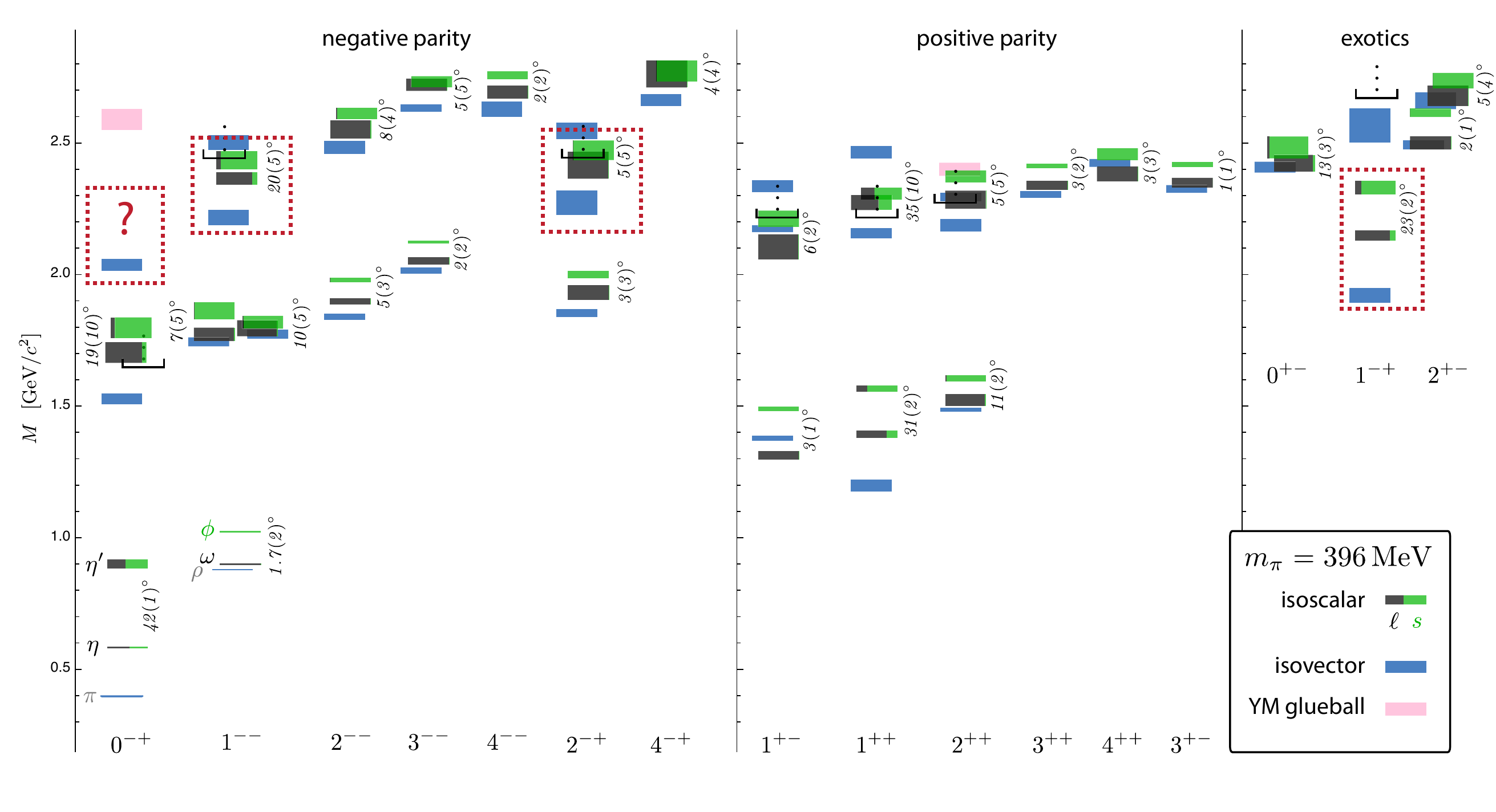}
\caption{\label{fig:lqcd_meson}A compilation of recent lattice QCD computations for both the
isoscalar and isovector light mesons from Ref.~\cite{Dudek:2011bn}, including 
$\ell\bar{\ell}$ $\left(|\ell\bar{\ell}\rangle\equiv (|u\bar{u}\rangle+|d\bar{d}\rangle)/\sqrt{2}\right)$ and
$s\bar{s}$ mixing angles (indicated in degrees).  The dynamical computation is
carried out with two flavors of quarks, light ($\ell$) and strange ($s$).  The $s$ quark
mass parameter is tuned to match physical $s\bar{s}$ masses, while the light quark mass parameters are
heavier, giving a pion mass of 396~MeV.  The black brackets with upward ellipses represent regions of the spectrum
where present techniques make it difficult to extract additional states.  
The dotted boxes indicate states that are interpreted as the lightest
hybrid multiplet -- the extraction of clear $0^{-+}$ states in this region is difficult in practice.}
\end{center}
\end{figure*}

As shown in Fig.~\ref{fig:lqcd_meson}, these calculations, for the first time, show a 
clear and detailed spectrum of exotic 
$J^{PC}$ mesons, with a lightest $1^{-+}$ lying a few hundred MeV below a $0^{+-}$ 
and two $2^{+-}$ states. Beyond this, through analysis of the matrix elements 
$\langle \mathfrak{n} | \mathcal{O}^\dag |0 \rangle$ for a range of different quark-gluon 
constructions, $\mathcal{O}$, we can infer \cite{Dudek:2011bn} that although the bulk of the 
non-exotic $J^{PC}$ spectrum has the expected systematics of a $q\bar{q}$ bound 
state system, some states are only interpolated strongly by operators featuring non-trivial 
gluonic constructions. One may interpret these states as non-exotic hybrid mesons, and, by combining them with
the spectrum of exotics, it is then possible to isolate a lightest hybrid supermultiplet of $(0,1,2)^{-+}$ and $1^{--}$ states, 
roughly 1.3 GeV heavier than the $\rho$ meson. The form of the operator that has strongest overlap 
onto these states has an $S$-wave $q\bar{q}$ pair in a color octet configuration and 
an exotic gluonic field in a color octet with $J_g^{P_gC_g}=1^{+-}$, a \emph{chromomagnetic} 
configuration. The heavier $(0,2)^{+-}$ states, along with some positive parity 
non-exotic states, appear to correspond to a $P$-wave coupling of the $q\bar{q}$ pair 
to the same chromomagnetic gluonic excitation.

A similar calculation for isoscalar states uses both $u\bar{u} + d\bar{d}$ and $s\bar{s}$ 
constructions and is able to extract both the spectrum of states and also their hidden 
flavor mixing.  (See Fig.~\ref{fig:lqcd_meson}.)  The basic experimental pattern of significant 
mixing in $0^{-+}$ and $1^{++}$ 
channels and small mixing elsewhere is reproduced, and, for the first time, we are able to say 
something about the degree of mixing for exotic-$J^{PC}$ states.  In order to
probe this mixing experimentally, it is essential to be able to reconstruct decays to both strange and non-strange
final state hadrons.

\begin{figure*}
\begin{center}
\includegraphics[width=0.8\linewidth]{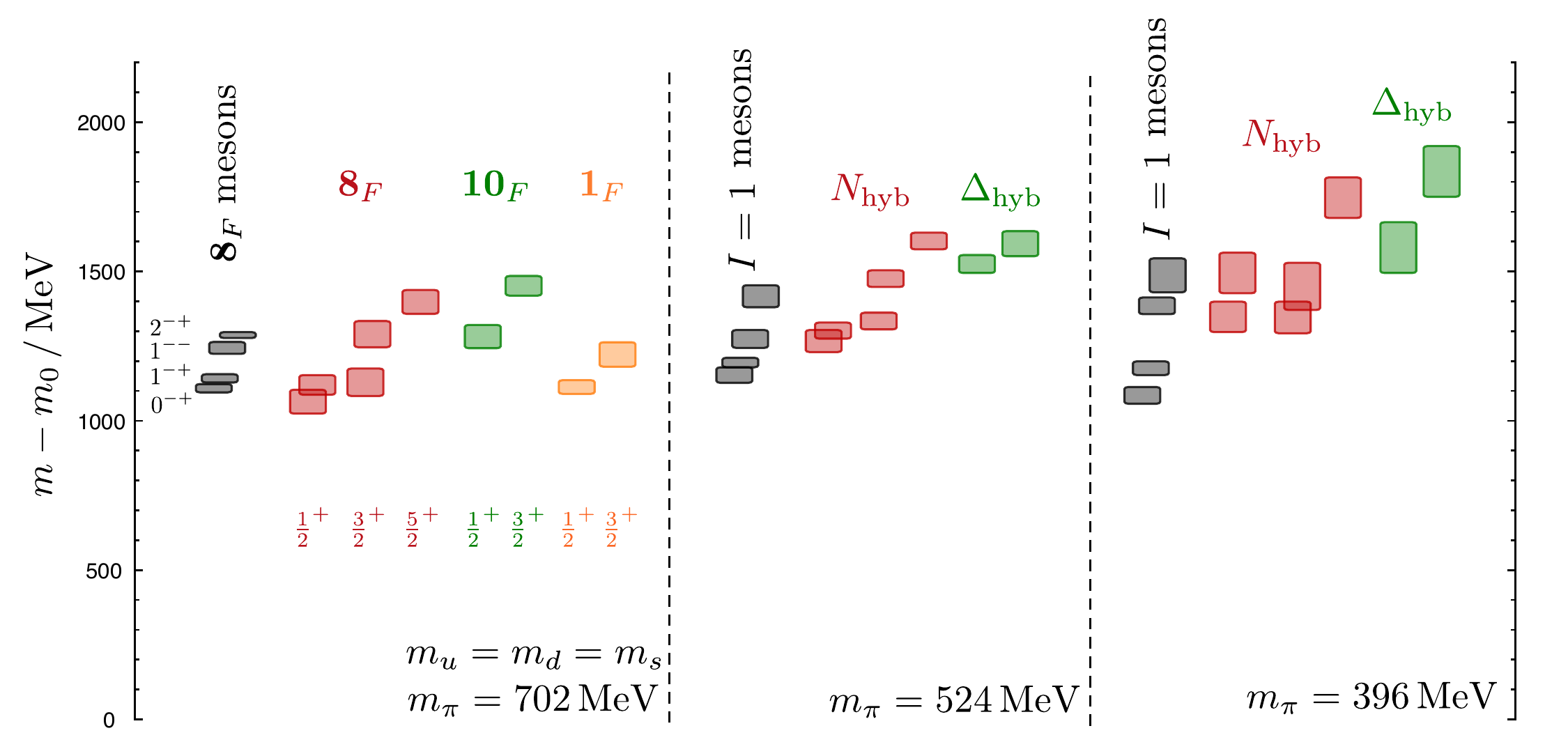}
\caption{\label{fig:baryon_meson} Spectrum of gluonic excitations in hybrid mesons (gray) and hybrid baryons (red, green, and orange) for three light quark masses.  The mass scale is $m-m_\rho$ for mesons and $m-m_N$ for baryons to approximately subtract the effect of differing numbers of quarks.  The left calculation is performed with perfect $SU(3)$-flavor symmetry, and hybrid members of the flavor octets ($8_F$), decuplet ($10_F$), and singlet ($1_F$) are shown.  The middle and right calculations are performed with a physical $s\bar{s}$ mass and two different values of $m_\pi$.}
\end{center}
\end{figure*}

A chromomagnetic gluonic excitation can also play a role in the spectrum of 
baryons:  constructions beyond the simple $qqq$ picture can occur when three quarks 
are placed in a color octet coupled to the chromomagnetic excitation. The 
baryon sector offers no ``smoking gun" signature for hybrid states, as all $J^P$ can be 
accessed by three quarks alone, but lattice calculations \cite{Edwards:2011jj} indicate that 
there are ``excess" nucleons with $J^P=1/2^+,~3/2^+,~5/2^+$ and excess $\Delta$'s 
with $J^P=1/2^+, 3/2^+$ that have a hybrid component. An interesting observation 
that follows from this study is that there appears to be a common energy cost for 
the chromomagnetic excitation, regardless of whether it is in a meson or baryon. 
In Fig.~\ref{fig:baryon_meson} we show the hybrid meson spectrum alongside the hybrid baryon 
spectrum with the quark mass contribution subtracted 
(approximately, by subtracting the $\rho$ mass from the mesons, and the nucleon mass from the baryons). 
We see that there appears to be a common scale $\sim 1.3$~GeV for the gluonic excitation, which does not 
vary significantly with varying quark mass.

Hybrid baryons will be challenging to extract experimentally because they lack ``exotic" character,
and can only manifest themselves by overpopulating the predicted spectrum with respect 
to a particular model. The current experimental situation of nucleon and $\Delta$~excitations is, 
however, quite contrary to the findings in the meson sector. Fewer baryon resonances are observed 
than are expected from models using three symmetric quark degrees of freedom, which does not 
encourage adding additional gluonic degrees of freedom. The current experimental efforts at 
Jefferson Lab aim to identify the relevant degrees of freedom which give rise to nucleon excitations.

Lattice calculations have made great progress at predicting the $N^\ast$~and $\Delta$~spectrum, 
including hybrid baryons~\cite{Edwards:2011jj,Dudek:2012ag}, and calculations are emerging for 
$\Xi$~and $\Omega$~resonances~\cite{Edwards:2012fx}. Experimentally, the properties of these multi-strange 
states are poorly known; only the $J^P$ of the $\Xi(1820)$ has been (directly) determined~\cite{Beringer:1900zz}. 
Previous experiments searching for Cascades were limited by low statistics and poor detector 
acceptance, making the interpretation of available data difficult. An experimental program on 
Cascade physics using the \gx~detector provides a new window of opportunity in hadron 
spectroscopy and serves as a complementary approach to the challenging study of broad and 
overlapping $N^\ast$~states. Furthermore, multi-strange baryons provide an important missing 
link between the light-flavor and the heavy-flavor baryons.

\subsection{Experimental context}
\label{sec:expcontext}

\gx~is ideally positioned to conduct a search for light-quark exotics and provide complementary
data on the spectrum of light-quark mesons.  It is anticipated that between now and
the time \gx~begins data taking, many results on the light-quark spectrum will have
emerged from the BESIII experiment, which is currently attempting to collect about $10^8$ to $10^9$
$J/\psi$ and $\psi'$ decays.  These charmonium states decay primarily through $c\bar{c}$
annihilation and subsequent  hadronization into light mesons, making them an ideal
place to study the spectrum of light mesons.  In fact several new states have already been 
reported by the BESIII collaboration such as the $X(1835)$, $X(2120)$, and $X(2370)$ in 
$J/\psi\to\gamma X$, $X\to\eta'\pi\pi$~\cite{Ablikim:2010au}.  No quantum number assignment
for these states has been made yet, so it is not yet clear where they fit into the meson spectrum.
\gx~can provide independent confirmation of the existence of these states in a completely different
production mode, in addition to measuring (or confirming) their $J^{PC}$ quantum numbers.
This will be essential for establishing the location of these states in the meson spectrum.
The BESIII experiment has the ability to reconstruct virtually any combination of final
state hadrons, and, due to the well-known initial state, kinematic fitting can be
used to effectively eliminate background. The list of putative new states and, therefore, the list of channels 
to explore with \gx, is expected to grow over the next few years as BESIII acquires and 
analyzes its large samples of charmonium data.

While the glue-rich $c\bar{c}$ decays of charmonium have long been hypothesized as
the ideal place to look for glueballs, decays of charmonium have also recently been used
to search for exotics.  The CLEO-c collaboration studied both $\pi^+\pi^-$ and
$\eta'\pi^\pm$ resonances in the decays of $\chi_{c1}\to\eta'\pi^+\pi^-$ and observed a
significant signal for an exotic $1^{-+}$ amplitude in the $\eta'\pi^\pm$ system~\cite{Adams:2011sq}.
The observation is consistent with the $\pi_1(1600)$ previously reported by E852 in
the $\eta'\pi$ system~\cite{Ivanov:2001rv}.  However, unlike E852, the CLEO-c analysis was 
unable to perform a model-independent extraction of the $\eta'\pi$ scattering amplitude
and phase to validate the resonant nature of the $1^{-+}$ amplitude.  A similar analysis
of $\chi_{c1}$ decays will most likely be performed by BESIII; however, even with an
order of magnitude more data, the final $\eta'\pi^+\pi^-$ sample is expected to be
just tens of thousands of events, significantly less than the proposed samples that will
be collected with \gx.  With the exception of this recent result from CLEO-c, the picture
in the light quark exotic sector, and the justification for building \gx, 
remains largely the same as it did at the time of the
original \gx~proposal; see Ref.~\cite{Meyer:2010ku} for a review.  All exotic candidates reported
to date are isovector $1^{-+}$ states ($\pi_1$).  By systematically exploring final states with both strange
and non-strange particles, \gx{} will attempt to establish not just one exotic state,
but a {\em pattern} of hybrid states with both exotic and non-exotic quantum numbers.

The idea that hybrids should also appear as supernumerary states in the spectrum of
non-exotic $J^{PC}$ mesons suggests an interesting interpretation of recent data in charmonium.
Three independent experiments have observed a state denoted $Y(4260)$~\cite{Aubert:2005rm,Coan:2006rv,He:2006kg,Yuan:2007sj};
it has $1^{--}$ quantum numbers but has no clear assignment in the arguably 
well-understood spectrum of $c\bar{c}$.  Even though the state is above the $D\bar{D}$ 
mass threshold, it does not decay strongly to $D\bar{D}$ as the other $1^{--}$ $c\bar{c}$ 
states in that region do.  Its mass is about 1.2~GeV above the ground state $J/\psi$,
which is similar to the splitting observed in lattice calculations of light mesons and baryons.
If this state is a non-exotic hybrid, an obvious, but very challenging, experimental goal would be to
identify the exotic $1^{-+}$ $c\bar{c}$ hybrid member of the same multiplet, which should have 
approximately the same mass\footnote{Like the light quark mesons discussed in Sec.~\ref{sec:theory}, 
the expectation for charmonium is that a $1^{--}$ non-exotic hybrid would exist with about the same mass
as the $1^{-+}$ exotic charmonium hybrid~\cite{Dudek:2008sz,Liu:2012ze}.}.  It is not clear how to produce such a state with existing experiments.  
In the light quark sector, some have suggested that the recently discovered $Y(2175)$~\cite{Aubert:2006bu,Ablikim:2007ab,Shen:2009zze} 
is the strangeonium ($s\bar{s}$) 
analogue of the $Y(4260)$.  If this is true, \gx~is well-positioned to study this state and
search for its exotic counterpart.  We discuss this further in Section~\ref{sec:normalss}.

Recent CLAS results~\cite{Price:2004xm,Guo:2007dw} also suggest many opportunities to make 
advances in baryon spectroscopy.  
The CLAS collaboration investigated $\Xi$ photoproduction in the reactions $\gamma p\to K^+K^+\,(X)$ as well as 
$\gamma p\to K^+K^+\pi^-\,(X)$ and, among other things, determined the mass splitting of the 
ground state $(\Xi^-,\,\Xi^0)$ doublet to be $5.4\pm 1.8$~MeV/$c^2$, which is consistent with 
previous measurements. Moreover, the differential cross sections for the production of the $\Xi^-$ have been 
determined in the photon energy range from 2.75 to 3.85~GeV~\cite{Guo:2007dw}. 
The cross section results are consistent with a production mechanism of $Y^\ast\to \Xi^-K^+$ 
through a $t$-channel process. The reaction $\gamma p\to K^+K^+\pi^-\,[\Xi^0]$ was also studied 
in search of excited $\Xi$ resonances, but no significant signal for an excited $\Xi$ state, 
other than the $\Xi^-(1530)$, was observed. The absence of higher-mass signals is very likely 
due to the low photon energies and the limited acceptance of the CLAS detector. 
With higher photon beam energy and two orders of magnitude more statistics, the \gx~experiment 
will be well-suited to search for and study these excited $\Xi$~resonances.

\section{Status of the \gx~experiment}


In the following section, we discuss the current status of the development of the 
baseline \gx~experiment.  The \gx~experiment was first presented to PAC~30 
in 2006~\cite{pac30}.   While beam time was not awarded for 12~GeV proposals 
at that PAC, the proposal suggested a three phase startup for \gx, which spanned 
approximately the first two calendar years of operation.  Phase I covered detector 
commissioning.  Phases II and III proposed a total of $7.5\times10^6$~s of 
detector live time at a flux of $10^7~\gamma$/s for physics commissioning and 
initial exploratory searches for hybrid mesons.  In 2010, an update of the experiment 
was presented to PAC~36 and a total of 120 days of beam time was granted for Phases I-III.  

In 2008, two critical detector components were ``de-scoped" from 
the design due to budgetary restrictions.  First, and most importantly, the forward 
Cherenkov particle identification system was removed.  The other component 
that was taken out was the level-three software trigger, which is needed for operating at a
photon flux greater than $10^7~\gamma$/s.  These changes severely impact the 
ultimate scientific goals and discovery potential of the \gx~experiment, as was 
noted in the PAC~36 report:

\begin{quote}
Finally, the PAC would like to express its hope that the de-scoped Cherenkov 
detector be revisited at some time in the future. The loss of kaon identification 
from the current design is a real shame, but entirely understandable given the 
inescapable limitations on manpower, resources, and time.
\end{quote}

In 2012, we proposed~\cite{pac39} both the implementation of the level-three trigger and the 
development of a kaon-identification system to be used during high intensity ($>10^7~\gamma$/s) 
running of \gx.  As noted in that proposal, the improved particle identification and the higher beam 
intensity would allow for a systematic exploration of higher-mass $s\bar{s}$ states as well as 
doubly-strange $\Xi$ baryons. In particular, identifying the $s\bar{s}$ members of the hybrid 
nonets and studying $s\bar{s}$ and $\ell\bar{\ell}$ 
$\left( |\ell\bar{\ell}\rangle\equiv (|u\bar{u}\rangle+|d\bar{d}\rangle)/\sqrt{2}\right)$ mixing 
amongst isoscalar mesons are crucial parts of the overall \gx{} program that would be
fully addressed by the PAC 39 proposal.  The PAC 39 proposal was conditionally approved, 
pending a final design of the kaon-identification system.

During the last twelve months, the collaboration has worked to better understand the
kaon identification capability of the baseline equipment, in addition to examining
how various additional particle identification detectors may augment this capability as
we work towards the goals of our PAC 39 proposal.
The level of detail with which we are now able to simulate the detector and
carry out complex analysis tasks has improved dramatically.  While we are still
converging on a hardware design for a kaon identification system, our studies have revealed that
an increase in intensity and running time alone is sufficient to {\em begin} a program
of studying mesons and baryons with hidden and open strangeness.  


\subsection{\gx~construction progress}

\begin{figure*}
\begin{center}
\includegraphics[width=0.8\linewidth]{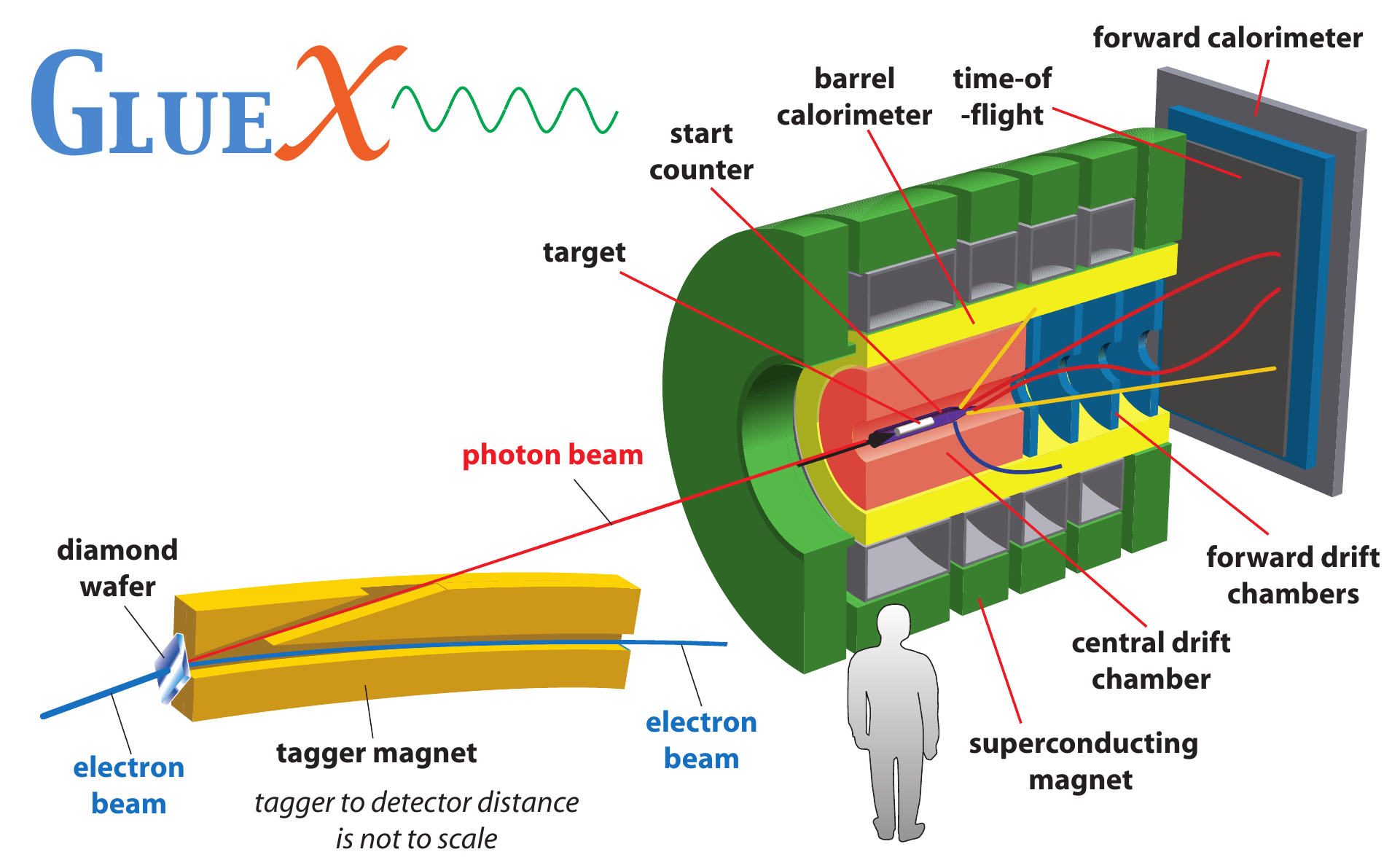}
\caption{\label{fig:detector}A schematic of the \gx~detector and beam.}
\end{center}
\end{figure*}

A schematic view of the \gx~detector is shown in Fig.~\ref{fig:detector}.  
The civil construction of Hall D is complete and the collaboration gained control of both
Hall~D and the Hall~D tagger hall in 2012. Many of the detector components are now being 
installed, with others being tested prior to installation. As of April 2013, all major sub-detector systems 
are either built or are under construction at Jefferson Lab or various collaborating 
institutions.  Beam for the experiment will be derived 
from coherent bremsstrahlung radiation from a thin diamond wafer and delivered to a liquid hydrogen 
target.  The solenoidal detector has both central and forward tracking chambers as well as central and 
forward calorimeters. Timing and triggering are aided by a forward time of flight wall and a thin scintillator 
start counter that encloses the target.  We briefly review the capabilities and construction status of each 
of the detector components below.  Table~\ref{tab:institutional_responsibilities} lists all of the 
\gx~collaborating experimental institutions and their primary responsibilities.  The collaboration consists of over a 
hundred members, including representation from the theory community.

\begin{table*}
\begin{center}
\caption{\label{tab:institutional_responsibilities}A summary of \gx~institutions and
their responsibilities. }
\begin{tabular}{l|l}\hline\hline
Institution & Responsibilities \\ \hline
Arizona State U. & beamline polarimetry, beamline support \\
Athens & BCAL and FCAL calibration \\
Carnegie Mellon U. & CDC, offline software, management \\
Catholic U. of America & tagger system \\
Christopher Newport U. & trigger system \\
U. of Connecticut & tagger microscope, diamond targets, offline software\\
Florida International U. & start counter \\
Florida State U. & TOF system, offline software \\
U. of Glasgow & goniometer, beamline support \\
Indiana U. & FCAL, offline software, management \\
Jefferson Lab & FDC, data acquisition, trigger, electronics, infrastructure, management \\
U. of Massachusetts & target, electronics testing \\
Massachusetts Institute of Technology& level-3 trigger, forward PID, offline software \\
MEPHI & offline and online software \\ 
Norfolk State U. & installation and commissioning \\
U. of North Carolina A\&T State & beamline support \\
U. of North Carolina, Wilmington & pair spectrometer \\
Northwestern U. & detector calibration \\
U. T\'ecnica Federico Santa Mar\'ia & BCAL readout \\
U. of Regina & BCAL, SiPM testing \\ \hline\hline
\end{tabular}
\end{center}
\end{table*}

\subsubsection{Beamline and Tagger}

The \gx~photon beam originates from coherent bremsstrahlung radiation produced by the 12~GeV electron beam
impinging on a $20~\mu$m diamond wafer.  Orientation of the diamond and downstream collimation
produce a photon beam peaked in energy around 9~GeV with about 40\% linear polarization.
A coarse tagger tags a broad range of electron energy, while precision tagging in the coherent peak 
is performed by a tagger microscope.  A downstream pair spectrometer is utilized to measure photon 
conversions and determine the beam flux.  Construction of the full system is underway.

Substantial work has also been done by the Connecticut group to fabricate and characterize thin 
diamond radiators for \gx. This has included collaborations with the Cornell High Energy Synchrotron 
Source as well as industrial partners.  Successful fabrication of 20~$\mu$m diamond 
radiators for \gx~has been demonstrated using a laser-ablation system at Connecticut. This system
starts with a much thicker diamond wafer and thins the central part of the diamond down to the
$20~\mu$m thickness, leaving a thicker picture frame around the outside of the diamond. This frame
allows for easier mounting and limits the vibration seen in thin diamonds. The design of the goniometer 
system to manipulate the diamond has been completed by the Glasgow group and the device has been
purchased from private industry.

The tagger magnet and vacuum vessel are currently being installed in the Hall D tagger 
hall. The design for the precision tagger ``microscope" was developed at Connecticut, including the 
custom electronics for silicon photomultiplier (SiPM) readout. Beam tests of prototypes have been 
conducted, and the construction of the final system is underway at Connecticut. The coarse tagger, 
which covers the entire energy range up to nearly the endpoint, is currently being built by the Catholic 
University group.

The groups from the University of North Carolina at Wilmington, North Carolina 
A\&T State, and Jefferson Lab are collaborating to construct the pair spectrometer.  A magnet obtained 
from Brookhaven has been modified to make it suitable for use in Hall D and is ready for installation.
In addition, the Arizona State and Glasgow groups are collaborating to develop a technique
for accurately measuring the linear polarization of the beam.  Tests are planned in Mainz this year.

\subsubsection{Solenoid~Magnet}
At the heart of the \gx~detector is the $2.2$~T superconducting solenoid, which provides
 the essential magnetic field for tracking.  The solenoidal geometry also has the benefit of reducing 
electromagnetic backgrounds in the detectors since low energy $e^+e^-$ pairs spiral within a small 
radius of the beamline. The field is provided by four superconducting coils.  These four coils have 
been tested independently with three of the four having been tested up to the nominal current of 
$1500$~A while the remaining coil was only tested to $1200$~A due to a problem with power leads 
that was unrelated to the coil itself.  No serious problems were found. The magnet has now been fully
assembled, and the solenoid has been operated at $1500$~A inside of Hall~D. At the time of submission
of this proposal, these studies are still underway.

\subsubsection{Tracking}

Charged particle tracking is performed by two systems:  a central straw-tube drift chamber (CDC) and 
four six-plane forward drift chamber (FDC) packages.  The CDC is composed of 28 layers of 1.5-m-long 
straw tubes.  The chamber provides $r-\phi$ measurements for charged tracks.  Sixteen of the 28 layers 
have a $6^\circ$ stereo angle to supply $z$ measurements.  Each FDC package is composed of six planes 
of anode wires.  The cathode strips on either side of the anode cross at $\pm75^\circ$ angles, providing
a two-dimensional intersection point on each plane.  The construction of the 
CDC~\cite{VanHaarlem:2010yq} has been completed by Carnegie Mellon University (CMU) and the 
chamber is currently being tested prior to delivery and installation in Hall D late in 2013. The 
construction of the FDC by Jefferson Lab is also complete, and  the chamber packages are undergoing 
testing prior to installation in Hall D in the fall of 2013.
The position resolution of the CDC and FDC is about $150~\mu$m and $200~\mu$m, respectively.  
Together the approximate momentum resolution is 2\%, averaged over the kinematical regions of 
interest.

Construction on the CDC began in May of 2010 with initial procurement and quality assurance of 
components and the construction of a 400~ft$^2$ class $2000$ cleanroom at CMU.
In August of that year, the end plates were mounted on the inner shell and then
aligned. The empty frame was then moved into a vertical position for the installation of
the 3522 straw tubes. This work started in November of 2010 and continued until October of 2011, 
when the outer shell was installed on the chamber.  Stringing of the wires was completed in 
February of 2012, and all tension voltage and continuity checks were completed in March of 2012.
In May of 2012, the upstream gas plenum was installed on the chamber and wiring of the high-voltage
system commenced.  This latter work finished
in early 2013 at which point the chamber was checked for gas tightness and the down-stream 
gas window was installed. 

After successful studies with a full-scale prototype, the FDC construction started 
in the beginning of 2011, with the entire production process carried out by Jefferson Lab in an 
off-site, 2000~ft$^2$ class 10,000 cleanroom. As of early 2013, all four packages had been completed and a spare
is being constructed using the extra parts. Tests of the packages are being carried out with cosmic rays 
in a system that uses external chambers for tracking, scintillators for triggering, and a DAQ system. 
With the anticipated delivery of the needed flash-ADC modules in 2013, full package readout tests
will be carried out.  The chamber is scheduled to be installed in the fall of 2013.

\subsubsection{Calorimetry}

Like tracking, the \gx~calorimetry system consists of two detectors:  a barrel calorimeter with a 
cylindrical geometry (BCAL) and a forward lead-glass calorimeter with a planar geometry (FCAL).  
The primary goal of these systems is to detect photons that can be used to reconstruct $\pi^0$'s and 
$\eta$'s, which are produced in the decays of heavier states.  The BCAL is a relatively high-resolution 
sampling calorimeter, based on 1~mm double-clad Kuraray scintillating fibers embedded in a lead matrix.
It is composed of 48 four-meter-long modules; each module having a radial thickness of 15.1 radiation lengths.
Modules are read out on each end by silicon SiPMs, which are not adversely 
affected by the high magnetic field in the proximity of the \gx~solenoid flux return.
The forward calorimeter is composed of 2800 lead glass modules, stacked in a circular array.  Each
bar is coupled to a conventional phototube.  The fractional energy resolution of the combined calorimetry
system $\delta(E)/E$ is approximately $5\%$-$6\%/\sqrt{E~[\mathrm{GeV}]}$.  Monitoring systems for both
detectors have been designed by the group from the University of Athens.

All $48$ BCAL calorimeter modules and a spare have been fabricated by the University of Regina
and are at Jefferson Lab where they have been fitted with light guides and sensors. These light guides 
have been fabricated at the University of Santa Mar\'ia (USM), which has also been responsible for testing 
of most of the Hamamatsu S12045X MPPC arrays (SiPMs). The LED calibration system for the BCAL has been
built by the University of Athens and has been installed as well. The assembled  modules are nearly ready 
for the start of their installation into the GlueX detector.

The 2800 lead glass modules needed for the FCAL have been assembled at Indiana University and shipped 
to Jefferson Lab. They are now stacked in the detector frame in Hall D, and work is proceeding on the 
remaining infrastructure and cabling to support the readout of the detector. All of the PMTs are  
powered by custom-built Cockcroft-Walton style photomultiplier bases~\cite{Brunner:1998fh} in order 
to reduce cable mass, 
power dissipation, and high voltage control system costs.  The design, fabrication, and testing of 
the bases was completed at Indiana University.
In addition, a 25-block array utilizing the production design of all FCAL components was 
constructed and tested with electrons in Hall~B by the Indiana University group in the spring of 2012; 
results indicate that the performance meets or exceeds expectations~\cite{FCALBeamTestNIM}.

\subsubsection{\label{sec:pid}Particle ID and timing}

The particle ID capabilities of \gx{} are derived from several subsystems.  A dedicated 
forward time-of-flight wall (TOF), which is constructed from two planes of 2.5-cm-thick scintillator 
bars, provides about $70$~ps timing resolution on forward-going tracks within about $10^\circ$ of 
the beam axis.  This information is complemented by time-of-flight data from the BCAL and specific
 ionization ($dE/dx$) measured with the CDC, both of which are particularly important for identifying 
the recoil proton in $\gamma p\to Xp$ reactions.  Finally, identification of the beam bunch, which 
is critical for timing measurements, is performed by a thin start counter that surrounds the target.

The TOF system is currently under construction at Florida State University. A prototype 
built using the final design hardware has achieved $100$~ps resolution for mean time from a single 
counter read-out from both ends. The system consists of two planes of such counters, implying
that the demonstrated two-plane resolution is $70$~ps. The detector is expected to be installed
in Hall D late in 2013.

Engineering drawings for the start counter are under development.  The counters and the 
electronics have to fit into a narrow space between the target vacuum chamber and the 
inner wall of the CDC.  Prototypes have obtained a 
time resolution of 300 to 600~ps, depending on the position 
of the hit along the length of the counter. The final segmentation has been fixed. 
SiPMs will be used for readout because they can be placed in the high 
magnetic field environment very close to the counters, thereby
preserving scintillation light. The design of the 
SiPM electronics is about to start, and a final prototype of the scintillator assembly is under development.

The combined PID system in the baseline design is sufficient for identification of most protons in the 
kinematic regions of interest for \gx.  The forward PID can be used to enhance the purity of the charged 
pion sample.  However, the combined momentum, path length, and timing resolution only allows for 
exclusive kaon identification for track momenta less than 2.0~GeV/$c$. However, 
because the hermetic \gx{} detector often reconstructs all final state particles, one can test
conservation of four-momentum via a kinematic fit as a means of particle identification.
This is especially effective when the recoil nucleon is measured. While it is true that no single particle 
identification measurement in \gx{} provides complete separation between kaons and pions, 
the contributions of many different but correlated measurements can provide effective PID.
 
\subsection{\label{sec:software}\gx{} software readiness}

Jefferson Lab organized an external review of the software efforts in all aspects of the 12~GeV
project in order to assess the status of software development in all experimental halls as
well as identify any issues that would impede progress toward the 12~GeV physics goals.
This review took place in June, 2012, and the report, issued in September~\cite{ReviewReport},
stated 

\begin{quote}
Overall, the Committee was very impressed with the current state of software and
computing preparations and the plans leading up to 12 GeV data taking.
\end{quote}

The sophistication of \gx{} simulation and reconstruction software was positively received by the committee.  The recommendations
of the committee focused on large scale implementation of these tools on both large data sets
and with large groups of people.  Recommendations were:
\begin{itemize}
\item The data volume and processing scale of GlueX is substantial but plans for data management 
and workload management systems supporting the operational scale were not made clear. They should
be carefully developed.
\item A series of scale tests ramping up using JLab's LQCD farm should be planned and conducted.
\end{itemize}

The \gx{} collaboration responded to both of these. To address the first recommendation and
prepare for the second recommendation several steps were taken.
\begin{itemize}
\item The format for reconstructed \gx{} data has been defined and implemented in
all \gx{} software. The size of the reconstructed data set is smaller than the estimates 
made at the time of the software review and lead us to believe that it should be possible to keep 
\gx{} data live on disk at multiple sites.
\item The format for raw data has been developed in collaboration with the data acquisition
group. The typical size of these events is about 50\% of what was originally estimated, but 
there remains significant uncertainty in how much this size may be inflated by detector noise.
\item To address analysis workload management, we have developed an analysis framework that allows
high-level, uniform access to a standardized set of reconstructed data as well as a consistent interface to 
analysis tasks such as kinematic fitting and particle identification code.  Our intent is to make
this a standard platform that facilitates easy analysis of any reaction by any member of the collaboration.
\end{itemize}

With this new infrastructure, the collaboration conducted a data challenge in December 2012, where 
over $5\times 10^{9}$ $\gamma p$ inclusive hadronic physics events were simulated and reconstructed. The point of this 
effort was not only to generate a large data sample for physics studies, but to also stress the robustness and
throughput capability of the current \gx{} software at scales comparable to production running, in line with the 
committee recommendation.  The five-billion event 
sample represents more than half of what will be collected in a year of Phase~III \gx{} running 
($10^{7}\, \gamma/s$). These events were simulated and reconstructed on a combination 
of the Open-science Grid (OSG) ($4\times 10^{9}$ events), the Jefferson Lab farm ($1\times 10^{9}$
events) and the Carnegie Mellon computer cluster ($0.3\times 10^{9}$ events).  We plan to incorporate
the lessons learned  from this data challenge into another similar exercise later this year.

It is this large data sample, coupled with smaller samples of specific
final states and new sophisticated analysis tools, that has allowed us to better understand 
the performance and capabilities of the baseline \gx{} detector.  
This ultimately has given us confidence that the baseline \gx{} detector is capable of carrying out 
initial studies of the $s\bar{s}$ hybrid spectrum for a select number of final states.  

\begin{figure*}
\begin{center}
\includegraphics[width=0.7\linewidth]{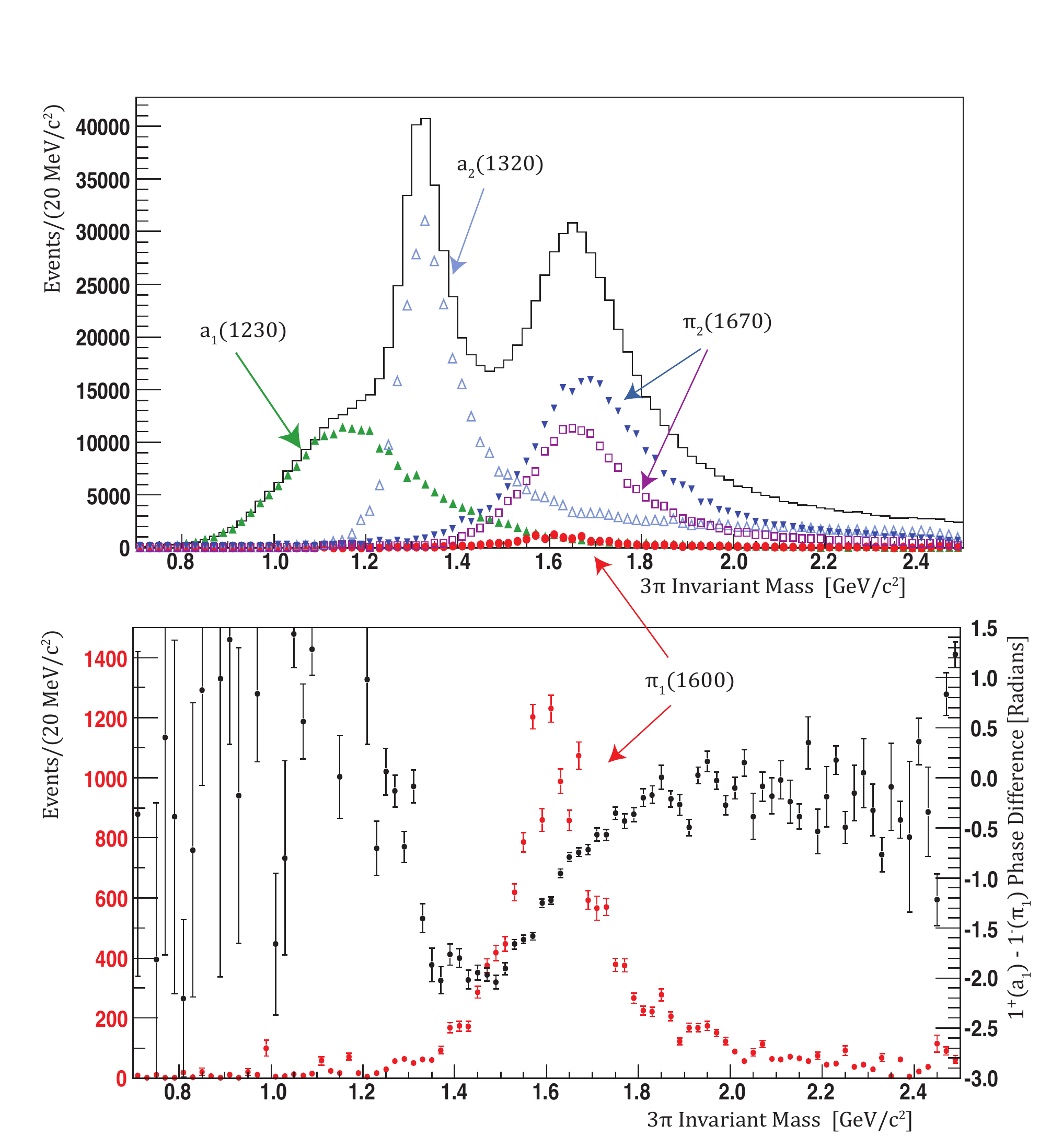}
\caption{\label{fig:amp_analysis} A sample amplitude analysis result for the 
$\gamma p \to \pi^+\pi^-\pi^+ n$ channel with \gx.  (top) The invariant mass spectrum as a function 
of $M(\pi^+\pi^-\pi^+)$ is shown by the solid histogram.  The results of the amplitude decomposition
 into resonant components in each bin is shown with points and error bars.  (bottom)  The exotic 
amplitude, generated at a relative strength of 1.6\%, is cleanly extracted (red points).  The black 
points show the phase between the $\pi_1$ and $a_1$ amplitudes.}
\end{center}
\end{figure*}

\subsection{Sensitivity and limitations to physics goals during initial \gx~running}

\subsubsection{Initial \gx{} physics goals:  non-strange final states}

Phases~I-III of the \gx~physics program provide an excellent opportunity for both the study of 
conventional mesons and the search for exotic mesons in photoproduction.
When one considers both production cross sections and detection efficiencies, the final states 
of interest will most likely focus on those decaying into non-strange mesons:  
$\pi$, $\eta$, $\eta^\prime$, and $\omega$.
Table~\ref{tab:exotic_modes} summarizes the expected lowest-mass exotics and possible decay modes.
Initial searches will likely focus on the $\pi_1$ isovector triplet and the $\eta_1$ isoscalar.  It will also be 
important to try to establish the other (non-exotic) members of the hybrid multiplet:  the $0^{-+}$, 
$1^{--}$, and $2^{-+}$ states.  Finally, the initial data may provide an opportunity to search for the 
heavier exotic $b_2$ and $h_2$ states. 

One reaction of interest is $\gamma p\to \pi^+\pi^-\pi^+ n$.  The $(3\pi)^\pm$ system has been 
studied extensively with data from E852~\cite{Adams:1998ff,Dzierba:2005jg} and 
COMPASS~\cite{Alekseev:2009aa}, with COMPASS reporting evidence 
for the exotic $\pi_1(1600)\to\rho\pi$ decay.  CLAS~\cite{Nozar:2008aa} has placed an upper limit
on the photoproduction of the $\pi_1(1600)$ and the subsequent decay to $\rho\pi$.  We have used
this limit as a benchmark to test the \gx{} sensitivity to small amplitudes by performing an amplitude analysis 
on a sample of purely generated 
$\gamma p\to \pi^+\pi^-\pi^+ n$ events that has been subjected to full detector simulation and 
reconstruction as discussed above.  Several conventional resonances, the $a_1$, $\pi_2$, and $a_2$, 
were generated along with a small ($<2\%$) component of the exotic $\pi_1$.  The result of the fit is 
shown in Figure~\ref{fig:amp_analysis}; the exotic amplitude and its phase are clearly extracted. 

\gx{} plans to systematically explore other non-strange channels, especially those that are predicted to be 
favorable hybrid decays.  One such study is to search for hybrid decays to $b_1\pi$, which,
considering the $b_1\to\omega\pi$ decay, results in a 5$\pi p$ final state.  In an analysis 
of mock data, we were able to use event selection and amplitude analysis to extract production of an exotic hybrid 
decay to $b_1\pi$ that is produced at a level corresponding to 0.03\% of the total hadronic 
cross section~\cite{IgorThesis}.  The signal to background ratio for the $\omega\pi\pi p$ sample exceeded
10:1.  Both of these studies indicate that the \gx~detector provides excellent sensitivity to 
exotic mesons that decay into non-strange final states.

As detailed later, production cross sections for many of these non-strange topologies of interest
are not well-known.  Data from pion production experiments or branching fractions of heavy mesons
suggest that production of $\eta$ and especially $\eta^\prime$ might be suppressed.  This implies
that a high-statistics study of, for example, $\gamma p \to \eta^\prime \pi^+ n$ to search for the
exotic state reported by E852~\cite{Ivanov:2001rv} or a study of the $f_1\pi$ final state, which populates
$\eta\pi\pi\pi$, will likely need the data derived from the high-intensity run put forth in this proposal.

\subsubsection{\label{sec:bdt} \gx~sensitivity to final states with strangeness}

\gx{} does not contain any single detector element that is capable of providing
discrimination of kaons from pions over the full-momentum range of interest for many
key reactions.  However, the hermetic \gx{} detector is capable of exclusively reconstructing all 
particles in the final state.  In the case where the recoil nucleon is a proton that is detectable
by the tracking chamber, this exclusive reconstruction becomes a particularly powerful tool for 
particle identification because conservation of four-momentum can be checked, via a kinematic
fit, for various mass hypotheses for the final state particles.  Many other detector quantities 
also give an indication of the particle mass, as assumptions about particle mass (pion or kaon)
affect interpretation of raw detector information.

An incomplete list of potentially discriminating quantities include:
\begin{itemize}
\item The confidence level (CL) from kinematic fitting that the event is consistent with the desired final
state.
\item The CL(s) from kinematic fitting that the event is consistent with some other final states.
\item The goodness of fit ($\chi^{2}$) of the primary vertex fit.
\item The goodness of fit ($\chi^{2}$) of each individual track fit.
\item The CL from the time-of-flight detector that a track is consistent with the particle mass.
\item The CL from the energy loss ($dE/dx$) that a track is consistent with the particle type.
\item The change in the goodness of fit ($\Delta \chi^{2}$) when a track is removed from the 
primary vertex fit.
\item Isolation tests for tracks and the detected showers in the calorimeter system.
\item The goodness of fit ($\chi^{2}$) of possible secondary vertex fits.
\item Flight-distance significance for particles such as $K_{S}$ and $\Lambda$ that lead to secondary
vertices.
\item The change in goodness of fit ($\Delta \chi^{2}$) when the decay products of a particle that 
produces a secondary vertex are removed from the primary vertex fit. 
\end{itemize}
The exact way that these are utilized depends on the particular analysis, but it is generally 
better to try and utilize as many of these as possible in a collective manner, rather than simply
placing strict criteria on any one of them. This means that we take advantage of 
correlations between variables in addition to the variables themselves.  One method of assembling
multiple correlated measurements into a single discrimination variable is a boosted decision tree
(BDT)~\cite{ref:bdt}.

Multivariate classifiers are playing an increasingly prominent role in particle physics.  The inclusion of 
BDTs and artificial neural networks (ANNs) is now commonplace in analysis selection criteria.  BDTs 
are now even used in software triggers~\cite{ref:lhcbhlt,ref:bbdt}.  
Traditionally, analyses have classified candidates using a set of variables, such as a kinematic fit confidence level, charged-particle time of flight, energy loss ($dE/dx$), {\em etc.}, where cuts are placed on each of the input variables to enhance the signal.  In a BDT analysis, however, cuts on individual variables are not used; instead, a single classifier is formed by combining the information from all of the input variables.

The first step in constructing a decision tree (DT) is to split the data into two subsamples (of unequal size) using the input variable which gives the largest separation between signal and background.  The variable to split on is determined by looping over all inputs.  As a result of this split, one subsample should be mostly background and the other mostly signal.  We now refer to these subsamples as branches. The process is repeated for these two branches, looping over all input variables to produce a second generation of four branches.  The process is repeated again on this new generation of branches, and then again as necessary until the tree is complete.  The final set of branches, referred to as {\em leaves}, is reached when one of the following occurs: a branch contains only signal or background events and so cannot be split any further; a branch has too few events to be split (a parameter specified by the grower of the tree); or the maximum number of total leaves has been reached (also a parameter specified by the grower).  Correlations are exploited by ensuring that all the input variables are included each time, including those used in previous branch divisions.

The above process is carried out using a training data sample that consists of events (possibly simulated) where it is known which class, signal or background, each event belongs to.  A single DT will overtrain and learn some fine-structure aspects of the data sample used for training which are due to statistical limitations of the data used and not aspects of the parent probability density function, {\em i.e.}, it will train on fluctuations.  To counter this, a series of classifiers is trained which greatly enhances the performance.  The training sample for each member of the series is augmented based on the performance of previous members.   Incorrectly classified events are assigned larger weights or boosted in importance; this technique is referred to as boosting.  The result is that each successive classifier is designed to improve the overall performance of the series in the regions where its predecessors have failed.  In the end, the members of the series are combined to produce a single powerful classifier: a BDT.  The performance is validated using an independent data sample, called a validation sample, that was not used in the training.  If the performance is found to be similar when using the training (where it is maximally biased) and validation (where it is unbiased) samples, then the BDT performance is predictable.  Practically, the output of the BDT is a single number for each event that tends towards one for signal-like events but tends towards negative one for background-like events.  Placing a requirement on the minimum value of this classifier, which incorporates all independent information input to the BDT, allows one to enhance the signal purity of a sample.  For a pedagogical description of BDTs, see Ref.~\cite{ref:roe}.

We illustrate the effectiveness of the BDT method by examining a reaction of the type
\begin{eqnarray*}
\gamma p & \to & p K^{+}K^{-} \pi^{+} \pi^{-},
\end{eqnarray*}
where we only consider the case where the recoil proton is reconstructed.  A missing
recoil nucleon reduces the number of constraints in the kinematic fit, and, consequently,
dramatically diminishes the capability of the fit to discriminate pions from kaons.
One can build a BDT for the reaction of interest, and look at
the efficiency of selecting true signal events as a function of the sample purity.  These studies do not include 
the efficiency of reconstructing the tracks in the detector,
but start at the point where a candidate event containing five charged tracks has been found.   Selection efficiencies
are given in Table~\ref{tab:bdt_efficiencies}.
When coupled with estimates of the detection efficiency, these data suggest
that it may be possible to have 90\% pure kaon samples with an overall
efficiency that is acceptable for analysis.  We are limited to events where
a recoil proton is detected, and, if we desire higher purity (lower background),
then the efficiency drops dramatically.

\begin{table}[h!]\centering
\caption[]{\label{tab:bdt_efficiencies}The selection efficiency of the BDT method for physics 
channel leading to a $K^{+}K^{-}\pi^{+}\pi^{-}p$ final state, assuming a particular signal purity is desired. 
Efficiency is selection efficiency only and does not include reconstruction efficiency.  The BDT 
method has been optimized using the simulation-determined \gx{} tracking resolution and 
a 50\% degraded tracking resolution to account for potential resolution simulation errors.}
\begin{tabular}{cccc}\hline\hline
~ & Tracking & Signal & Selection  \\
Final State & Resolution & Purity & Efficiency \\ \hline
$K^{+}K^{-}\pi^{+}\pi^{-}p$ & Nominal & $0.90$ & $0.26$ \\
$K^{+}K^{-}\pi^{+}\pi^{-}p$ & Nominal & $0.95$ & $0.09$ \\
$K^{+}K^{-}\pi^{+}\pi^{-}p$ & Degraded  & $0.90$ & $0.25$ \\
$K^{+}K^{-}\pi^{+}\pi^{-}p$ & Degraded  & $0.95$ & $0.09$ \\
\hline\hline
\end{tabular}
\end{table}

\subsubsection{Limitations of existing kaon identification algorithms}

It is important to point out that the use of kinematic constraints 
to achieve kaon identification, without dedicated hardware, has limitations.  
By requiring that the recoil proton be reconstructed,
we are unable to study charge exchange processes that have a recoil neutron.  In addition,
this requirement results in a loss of efficiency of 30\%-50\% for proton recoil topologies and 
biases the event selection to those that have high momentum transfer,
which may make it challenging to conduct studies of the production mechanism.
Our studies indicate that it will be difficult to attain very high purity samples with
a multivariate analysis alone.  In channels with large cross sections, the \gx{}
sensitivity will not be limited by acceptance or efficiency, but by the ability to
suppress and parameterize backgrounds in the amplitude analysis.  To push the
limits of sensitivity we need not only high statistics but high purity.
Finally, it is worth noting that our estimates of the kaon selection efficiency
using kinematic constraints depends strongly on our ability to model the performance of the
detector.  Although we have constructed a complete simulation, the experience of
the collaboration with comparable detector systems indicates that the simulated
performance is often better than the actual performance in unforeseen ways.

While the studies of kaon-identification systems as part of the PAC 39 proposal
are not yet complete, some general comments can be made on the impact of additional kaon-identification.  
The obvious advantage of supplemental particle identification hardware is
that it provides new, independent information to the multivariate analysis that has a very
high discrimination power.  We see noticeable improvements in efficiency when information from
various design supplemental kaon ID systems is included in the BDT; this is especially dramatic
at 95\% purity.  

Despite the limitations noted above, we will demonstrate that a program of high
intensity running with the \gx~detector, even without a dedicated particle identification upgrade,
is capable of producing interesting results in the study of $s\bar{s}$ mesons and $\Xi$ baryons.
In addition, the order-of-magnitude increase in statistical precision in non-strange channels will allow us to
study production mechanisms ($t$-dependence of reactions) with greater precision and
to search for rarely produced resonances.

\section{\boldmath Study of $s\bar{s}$ Mesons}
\label{sec:ss_meson}

The primary goal of the \gx~experiment is to conduct a definitive mapping of states in the light 
meson sector, with an emphasis on searching for exotic mesons.  Ideally, we would like to produce 
the experimental analogue of the lattice QCD spectrum pictured in Fig.~\ref{fig:lqcd_meson}, enabling 
a direct test of our understanding of gluonic excitations in QCD.  In order to achieve this, one must be 
able to reconstruct strange final states, as observing decay patterns of mesons has been one of the 
primary mechanisms of inferring quark flavor content.  An example of this can be seen by examining 
the two lightest isoscalar $2^{++}$ mesons in the lattice QCD calculation in Fig.~\ref{fig:lqcd_meson}.  
The two states have nearly pure flavors, with only a small ($11^\circ$) mixing in the $\ell\bar{\ell}$ and 
$s\bar{s}$ basis.  A natural experimental assignment for these two states are the $f_2(1270)$ and the 
$f_2'(1525)$.  An experimental study of decay patterns shows that 
$\mathcal{B}(f_2(1270)\to K K)/\mathcal{B}(f_2(1270)\to \pi\pi)\approx 0.05$ and 
$\mathcal{B}(f_2'(1525)\to \pi\pi)/\mathcal{B}(f_2'(1525)\to K K) \approx 0.009$~\cite{Beringer:1900zz}, 
which support the prediction of an $f_2(1270)$ ($f_2'(1525)$) with a dominant $\ell\bar{\ell}$ ($s\bar{s}$) 
component.  By studying both strange and non-strange decay modes of mesons, \gx~hopes to provide
similarly valuable experimental data to aid in the interpretation of the hybrid spectrum.

\subsection{\boldmath Exotic $s\bar{s}$ states}

While most experimental efforts to date have focused on the lightest isovector exotic meson, the 
$J^{PC}=1^{-+}$ $\pi_1(1600)$, lattice QCD clearly predicts a rich spectrum of both isovector and
isoscalar exotics, the latter of which may have mixed $\ell\bar{\ell}$ and $s\bar{s}$ flavor content.
A compilation of the ``ground state" exotic hybrids is listed in Table~\ref{tab:exotic_modes}, along with
theoretical estimates for masses, widths, and key decay modes.  It is expected that initial searches 
with the baseline \gx~hardware will target primarily the $\pi_1$ state.  Searches for the $\eta_1$, $h_0$, and $b_2$ may be 
statistically challenging, depending on the masses of these states and the production cross sections.
With increased statistics and kaon identification, the search scope can be broadened to include these
heavier exotic states in addition to the $s\bar{s}$ states:  $\eta_1'$, $h_0'$, and $h_2'$.  The $\eta_1'$ and
$h_2'$ are particularly interesting because some models predict these states to be relatively narrow, and
that they should decay through well-established kaon resonances.

The observation of various $\pi_1$ states has been reported in the literature for over fifteen
years, with some analyses based on millions of events.  However, it is safe to say that there exists
a fair amount of skepticism regarding the assertion that unambiguous experimental evidence
exists for exotic hybrid mesons.  If the scope of exotic searches with \gx~is narrowed to only include
the lightest isovector $\pi_1$ state, the ability for \gx~to comprehensively address the question of 
the existence of gluonic excitations in QCD is greatly diminished.  On the other hand, clear identification of 
all exotic members of the lightest hybrid multiplet, the three exotic $\pi_1^{\pm,0}$ states and the exotic 
$\eta_1$ and $\eta_1'$, which can only be done by systematically studying a large number of
strange and non-strange decay modes, would provide unambiguous experimental confirmation of 
exotic mesons. A study of decays to kaon final states could demonstrate that the $\eta_1$ candidate 
is dominantly $\ell\bar{\ell}$ while the $\eta_1'$ candidate is $s\bar{s}$, as predicted by initial lattice 
QCD calculations.  Such a discovery would represent a substantial improvement in the experimental 
understanding of exotics.  In addition, further identification of members of the 
$0^{+-}$ and $2^{+-}$ nonets as well as measuring the mass splittings with the $1^{+-}$ states will 
validate the lattice QCD inspired phenomenological picture of these states as $P$-wave 
couplings of a gluonic field with a color-octet $q\bar{q}$ system.

\begin{table*}\centering
\caption{\label{tab:exotic_modes}
A compilation of exotic quantum number hybrid approximate masses, widths, and decay predictions.
Masses are estimated from dynamical LQCD calculations with $M_\pi = 396~\mathrm{MeV}/c^2$~\cite{Dudek:2011bn}.  
The PSS (Page, Swanson and Szczepaniak) and IKP (Isgur, Kokoski and Paton) model widths are from Ref.~\cite{Page:1998gz},
with the IKP calculation based on the model in Ref.~\cite{Isgur:1985vy}.  The total widths have a mass 
dependence, and Ref.~\cite{Page:1998gz} uses somewhat different mass values than suggested by the most recent
lattice calculations~\cite{Dudek:2011bn}.
Those final states marked with a dagger ($\dagger$) are ideal for experimental exploration 
because there are relatively few stable particles in the final state or moderately narrow 
intermediate resonances that may reduce combinatoric background.  
(We consider $\eta$, $\eta^\prime$, and $\omega$ to be stable final state particles.)}
\begin{tabular}{ccccccc}\hline\hline
 &  Approximate & $J^{PC}$ & \multicolumn{2}{c}{Total Width (MeV)} & 
Relevant Decays & Final States \\ 
         & Mass (MeV) &  & PSS & IKP & &  \\ \hline
$\pi_{1}$   & 1900 & $1^{-+}$ &  $80-170$ & $120$ & 
$b_{1}\pi^\dagger$, $\rho\pi^\dagger$, $f_{1}\pi^\dagger$, $a_{1}\eta$, $\eta^\prime\pi^\dagger$ & $\omega\pi\pi^\dagger$, $3\pi^\dagger$, $5\pi$, $\eta 3\pi^\dagger$, $\eta^\prime\pi^\dagger$  \\
$\eta_{1}$  & 2100 & $1^{-+}$ &  $60-160$ & $110$ &
$a_{1}\pi$, $f_{1}\eta^\dagger$, $\pi(1300)\pi$ & $4\pi$, $\eta 4\pi$, $\eta\eta\pi\pi^\dagger$ \\ 
$\eta^{\prime}_{1}$ & 2300 & $1^{-+}$ &  $100-220$ & $170$ &
$K_{1}(1400)K^\dagger$, $K_{1}(1270)K^\dagger$, $K^{*}K^\dagger$ & $KK\pi\pi^\dagger$, $KK\pi^\dagger$, $KK\omega^\dagger$ \\ \hline
$b_{0}$     & 2400 & $0^{+-}$ & $250-430$ & $670$ &
$\pi(1300)\pi$, $h_{1}\pi$ & $4\pi$ \\
$h_{0}$     &  2400 & $0^{+-}$ & $60-260$  & $90$  &
$b_{1}\pi^\dagger$, $h_{1}\eta$, $K(1460)K$ & $\omega\pi\pi^\dagger$, $\eta3\pi$, $KK\pi\pi$ \\
$h^{\prime}_{0}$    & 2500& $0^{+-}$ & $260-490$ & $430$ &
$K(1460)K$, $K_{1}(1270)K^\dagger$, $h_{1}\eta$ & $KK\pi\pi^\dagger$, $\eta3\pi$ \\ \hline
$b_{2}$     & 2500 & $2^{+-}$ &    $10$ & $250$ &
$a_{2}\pi^\dagger$, $a_{1}\pi$, $h_{1}\pi$ & $4\pi$, $\eta\pi\pi^\dagger$ \\
$h_{2}$     & 2500 & $2^{+-}$ &    $10$ & $170$ &
$b_{1}\pi^\dagger$, $\rho\pi^\dagger$ & $\omega\pi\pi^\dagger$, $3\pi^\dagger$ \\
$h^{\prime}_{2}$  & 2600 & $2^{+-}$ &    $10-20$ &  $80$ &
$K_{1}(1400)K^\dagger$, $K_{1}(1270)K^\dagger$, $K^{*}_{2}K^\dagger$ & $KK\pi\pi^\dagger$, $KK\pi^\dagger$\\
\hline\hline
\end{tabular}
\end{table*}

\subsection{\boldmath Non-exotic $s\bar{s}$ mesons}
\label{sec:normalss}

As discussed in Section~\ref{sec:theory}, one expects the lowest-mass hybrid multiplet to contain 
$(0,1,2)^{-+}$ states and a $1^{--}$ state that all have about the same mass and correspond to an 
$S$-wave $q\bar{q}$ pair coupling to the gluonic field in a $P$-wave.  For each $J^{PC}$ we 
expect an isovector triplet and a pair of isoscalar states in the spectrum.  Of the four sets of $J^{PC}$
values for the lightest hybrids, only the $1^{-+}$ is exotic. The other hybrid states will appear as 
supernumerary states in the spectrum of conventional mesons. The ability to clearly identify these 
states depends on having a thorough and complete understanding of the meson spectrum.  Like 
searching for exotics, a complete mapping of the spectrum of non-exotic mesons requires the 
ability to systematically study many strange and non-strange final states.  Other experiments, such 
as BESIII or COMPASS, are carefully studying this with very high statistics data samples and
and have outstanding capability to cleanly study any possible final state.  While the production 
mechanism of \gx~is complementary to that of charmonium decay or pion beam production and is 
thought to enhance hybrid production, it is essential that the detector capability and statistical 
precision of the data set be competitive with other contemporary experiments in order to maximize the collective
experimental knowledge of the meson spectrum. 

Given the recent developments in charmonium (briefly discussed in Section~\ref{sec:expcontext}),
a state that has attracted a lot of attention in the $s\bar{s}$ spectrum is the $Y(2175)$, which is
assumed to be an $s\bar{s}$ vector meson ($1^{--}$).  The $Y(2175)$ has been observed to decay to 
$\pi\pi\phi$ and has been produced in both $J/\psi$ decays~\cite{Ablikim:2007ab} and $e^+e^-$ 
collisions~\cite{Aubert:2006bu,Shen:2009zze}. The state is a proposed analogue of the $Y(4260)$ in 
charmonium, a state that is also about 1.2 GeV heavier than the ground state triplet ($J/\psi$)
and has a similar decay mode:  $Y(4260)\to\pi\pi J/\psi$.  The $Y(4260)$ has no obvious interpretation
in the charmonium spectrum and has been speculated to be a hybrid 
meson~\cite{Close:2005iz,Zhu:2005hp,Kou:2005gt,Luo:2005zg}, which, by loose analogy,
leads to the implication that the $Y(2175)$ might also be a hybrid candidate.  It should be noted
that the spectrum of $1^{--}$ $s\bar{s}$ mesons is not as well-defined experimentally as the $c\bar{c}$ system; 
therefore, it is not clear that the $Y(2175)$ is a supernumerary state.  However, \gx~is ideally suited
to study this system.  We know that vector mesons are copiously produced in photoproduction; therefore,
with the ability to identify kaons, a precision study of the $1^{--}$ $s\bar{s}$ spectrum can be conducted
with \gx.  Some have predicted~\cite{Ding:2007pc} that the potential hybrid nature of the $Y(2175)$ can be explored by studying ratios of branching fractions into various kaonic final states.  
In addition, should \gx~be able to conclude that the $Y(2175)$ is in fact a supernumerary 
vector meson, then a search can be made for the exotic $1^{-+}$ $s\bar{s}$ member of the multiplet 
($\eta_1'$),  evidence of which would provide a definitive interpretation of the $Y(2175)$ and likely have
implications on how one interprets charmonium data.

\subsection{\boldmath \gx~sensitivity to $s\bar{s}$ mesons}

Recent studies of the capability of the baseline \gx{} detector indicate that we will
have adequate sensitivity to a number of final states containing kaons. Generically, these appear
to be final states in which the recoil proton is reconstructed, as this provides the kinematic fit
with the most power to discriminate among particle mass hypotheses.  We discuss below
the \gx~sensitivity to a variety of final state topologies motivated by the physics topics in the
preceding sections.

Table~\ref{tab:exotic_modes} provides information from models of hybrid mesons
on the expected decay modes of exotic quantum-number states.
The $\eta_{1}^{\prime}$, the $h_{0}^{\prime}$, and the $h_{2}^{\prime}$ all couple to the $KK\pi\pi$ 
final state, while both the $\eta_{1}^{\prime}$ and the $h_{2}^{\prime}$ are expected to couple to the
$KK\pi$ final state. To study the \gx{} sensitivity to these two final states, we have modeled
two decay chains. For the $KK\pi$ state, we assume one of the kaons is a $K_{S}$, which leads to a secondary 
vertex and the $K^{+}\pi^{-}\pi^{+}\pi^{-}$ final state:
\begin{eqnarray}
\label{eq:eta1}
\eta_{1}^{\prime}(2300) & \rightarrow & K^{*}K_{S} \nonumber \\
                                  & \rightarrow & (K^{+}\pi^{-})(\pi^{+}\pi^{-}) \nonumber \\ 
                                  & \rightarrow & K^{+}\pi^{-}\pi^{+}\pi^{-}. 
\end{eqnarray}
For the $KK\pi\pi$ state we assume no secondary vertex:
\begin{eqnarray}
\label{eq:h2}
h_{2}^{\prime}(2600)      & \rightarrow & K_{1}^{+}K^{-} \nonumber \\
                                  & \rightarrow & (K^{*}(892)\pi^{+})K^{-} \nonumber \\
                                  & \rightarrow & K^{+}K^{-}\pi^{-}\pi^{+}. 
\end{eqnarray}

In addition to the exotic hybrid channels, there is an interest in non-exotic $s\bar{s}$ mesons. 
In order to study the sensitivity to conventional $s\bar{s}$ states, we consider 
an excitation of the normal $\phi$ meson, the known $\phi_{3}(1850)$, which decays to $K\bar{K}$ 
\begin{eqnarray}
\label{eq:phi1850}
\phi_{3}(1850) & \rightarrow & K^{+}K^{-} \, .
\end{eqnarray}
The detection efficiency of this state will be typical of $\phi$-like states decaying to the same
final state.

Finally, as noted in Section~\ref{sec:normalss}, the $Y(2175)$ state is viewed 
as a potential candidate for a non-exotic hybrid and has been reported in the decay mode
\begin{eqnarray}
\label{eq:Y2175}
Y(2175) & \rightarrow & \phi f_{0}(980) \nonumber \\
             & \rightarrow & (K^{+}K^{-})(\pi^{+}\pi^{-})  \, .
\end{eqnarray}
While this is the same $KK\pi\pi$ state noted in reaction~\ref{eq:h2} above, the intermediate resonances make
the kinematics of the final state particles different from the exotic decay channel noted above.
Therefore, we simulate it explicitly.

\begin{figure*}
\begin{center}
\includegraphics[width=\linewidth]{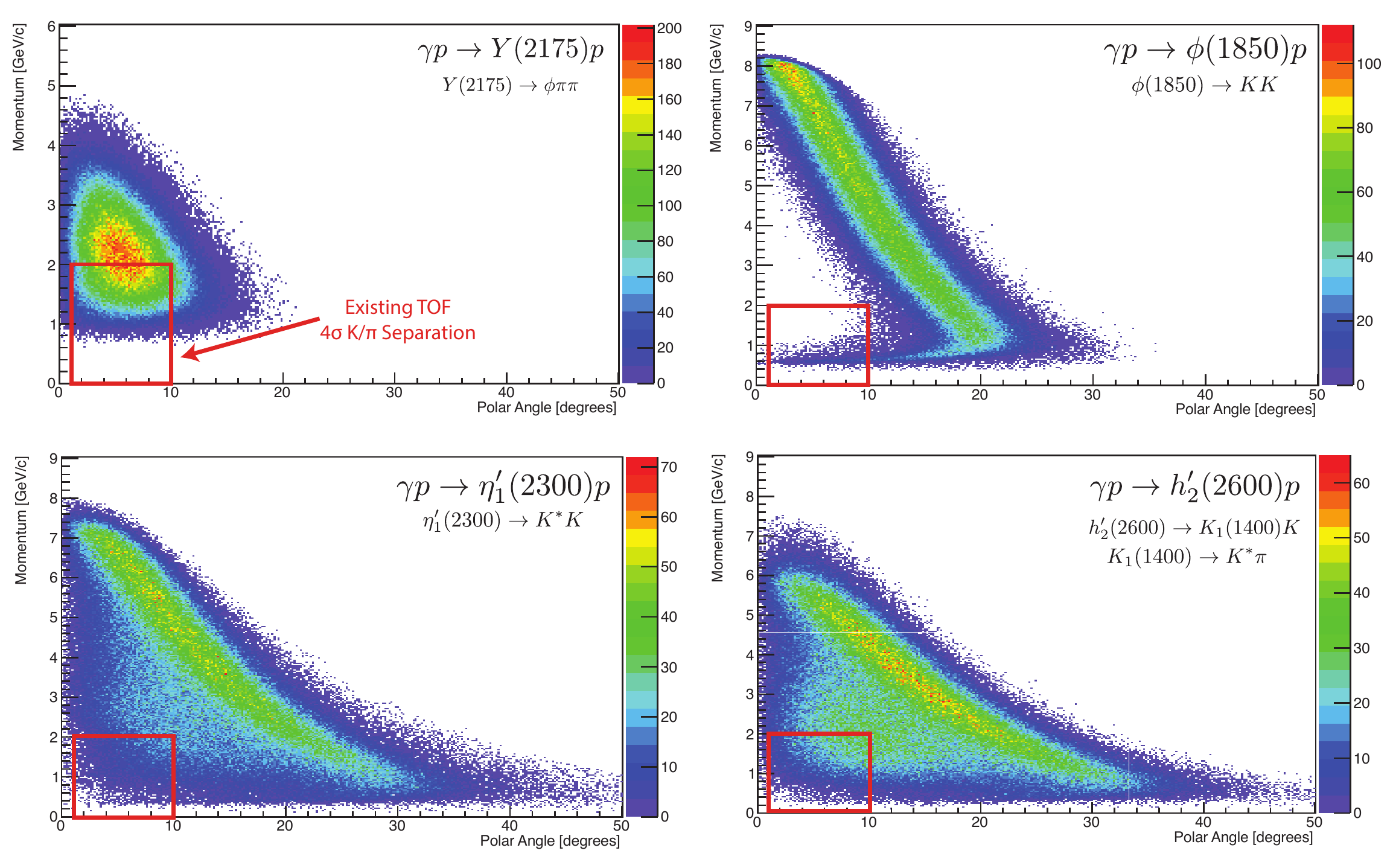}
\caption{\label{fig:kaon_kin}
Plots of particle density as a function of momentum and polar angle for all kaons in a variety of different 
production channels.  Shown in solid (red) is the region of phase space where the existing time-of-flight\ (TOF) detector in \gx{} provides $K/\pi$ discrimination at the four standard deviation level.}
\end{center}
\end{figure*}

The final-state kaons from the reactions~\ref{eq:eta1} - \ref{eq:Y2175}
will populate the \gx{} detectors differently, with different overlap of the region
where the time-of-flight system can provide good $K$/$\pi$ separation. Figure~\ref{fig:kaon_kin} 
shows the kinematics of these kaons and the overlap with the existing time-of-flight sensitivity.

A BDT analysis (see Section~\ref{sec:bdt}) has been used to study the capabilities of the baseline 
\gx{} detector to identify the four reactions of interest.  This
study used the \textsc{pythia}-simulated $\gamma p$ collisions from the large-scale data challenge as described in
Section~\ref{sec:software}.  Signal samples were obtained from \textsc{pythia} events with the generated 
reaction topology, and the remainder of the inclusive photoproduction reactions were used as the 
background sample. A large number of discriminating variables were used in the BDT analysis,
which generated a single classifier by combining the information from all of the input variables. (The BDT 
algorithms used are contained within ROOT in the TMVA package~\cite{ref:TMVA2007}.) 

In all cases we set the requirement on the BDT classifier in order to obtain a fixed final sample
purity.  For example, a purity of 90\% implies a background at the 10\% level.  Any exotic signal
in the spectrum would likely need to be larger than this background to be robust.  Therefore, with
increased purity we have increased sensitivity to smaller signals, but also lower efficiency.  
In Table~\ref{tab:bdt_eff}, we present the signal selection efficiencies (post reconstruction)
for our four reactions of interest assuming the design resolution of the \gx{} tracking system. As noted
earlier, these assume that the tracks have been reconstructed and do not include that efficiency. 
Historical evidence suggests that simulated resolutions are always more optimistic than what
is attainable with the actual detector.  To check the sensitivity of our conclusions to such
a systematic error, we repeat the study while degrading the tracking resolution by 50\%.  At the 90\%
purity level, this degradation reduces the efficiency noticeably but not severely.
\begin{table}[h!]\centering
\caption[]{\label{tab:bdt_eff}Efficiencies for identifying several final states in \gx . The efficiencies do 
not include the reconstruction of the final state tracks.}
\begin{tabular}{cccc} \hline\hline
~ & Tracking & Signal & Selection \\
Meson Decay & Resolution & Purity & Efficiency \\ \hline
$\phi_{3}(1850)\to K^+K^-$   & Nominal & $0.90$ & $0.73$ \\
$Y(2175)\to \phi f_0(980)$   & Nominal & $0.90$ & $0.53$ \\ 
$\eta_{1}^{\prime}(2300)\to K^*K_S$   & Nominal & $0.90$ & $0.32$ \\
$h_{2}^{\prime}(2600)\to K_1^+K^-$   & Nominal & $0.90$ & $0.26$ \\ \hline
$\phi_{3}(1850)\to K^+K^-$   & Degraded & $0.90$ & $0.73$ \\
$Y(2175)\to \phi f_0(980)$   & Degraded & $0.90$ & $0.49$ \\ 
$\eta_{1}^{\prime}(2300)\to K^*K_S$   & Degraded  & $0.90$ & $0.31$ \\
$h_{2}^{\prime}(2600)\to K_1^+K^-$   & Degraded  & $0.90$ & $0.25$ \\ \hline
$\phi_{3}(1850)\to K^+K^-$   & Nominal & $0.95$ & $0.67$ \\
$Y(2175)\to \phi f_0(980)$   & Nominal & $0.95$ & $0.31$ \\ 
$\eta_{1}^{\prime}(2300)\to K^*K_S$   & Nominal & $0.95$ & $0.15$ \\
$h_{2}^{\prime}(2600)\to K_1^+K^-$   & Nominal & $0.95$ & $0.09$ \\
\hline\hline
\end{tabular}
\end{table}

Finally, we have studied the resulting efficiency when we require a signal purity of 95\%,
which, for example, would be necessary to search for more rare final states.  
As can be seen in Table~\ref{tab:bdt_eff}, increasing the desired purity noticeably reduces the
efficiency:  in two of the four topologies studied the efficiency drops by over 50\% of itself as the
desired purity is increased from 90\% to 95\%.  This exposes the limit
of what can be done with the baseline \gx{}~hardware.  Preliminary studies with
supplemental kaon identification hardware (similar to those discussed in our PAC 39 proposal)
indicate that very high-purity samples are attainable with significantly improved efficiency.
It is likely that studies of final states where the background must be reduced below 10\%
will need additional particle identification hardware.

\section{\boldmath $\Xi$ baryons}
\label{sec:xi_baryon}

The spectrum of multi-strange hyperons is poorly known, with only a
few well-established resonances. Among the
doubly-strange states, the two ground-state $\Xi$'s, the octet
member~$\Xi(1320)$ and the decuplet member $\Xi(1530)$, have four-star
status in the PDG~\cite{Beringer:1900zz}, with only four other
three-star candidates. On the other hand, more than 20~$N^\ast$ and
$\Delta^\ast$ resonances are rated with at least three stars in the
PDG. Of the six $\Xi$~states that have at least a three-star rating in
the PDG, only two are listed with weak experimental evidence for their
spin-parity $(J^P)$ quantum numbers: $\Xi(1530)\frac{3}{2}^+$~\cite{Aubert:2008ty} and
$\Xi(1820)\frac{3}{2}^-$~\cite{Biagi:1986vs}. All other $J^P$~assignments,
including the $J^P$ for the $\Xi(1320)$ ground state, 
are based on quark-model predictions. Flavor~$SU(3)$ symmetry predicts
as many $\Xi$~resonances as $N^\ast$ and $\Delta^\ast$~states
combined, suggesting that many more $\Xi$ resonances remain
undiscovered. The three lightest quarks, $u$, $d$, and $s$, have 27
possible flavor combinations: $3\otimes 3\otimes 3 = 1\oplus 8\oplus
8\,^{\prime}\oplus 10$ and each multiplet is identified by its spin
and parity,~$J^P$. Flavor $SU(3)$ symmetry implies that the members of
the multiplets differ only in their quark makeup, and that the basic
properties of the baryons should be similar, although the symmetry is
known to be broken by the strange-light quark mass difference.
The octets consist of $N^*$, $\Lambda^*$, $\Sigma^*$, and
$\Xi^*$~states. We thus expect that for every $N^*$ state, there
should be a corresponding $\Xi^*$~state \emph{with similar
  properties}. Additionally, since the decuplets consist of
$\Delta^*$, $\Sigma^*$, $\Xi^*$, and  $\Omega^*$~states, we also
expect for every $\Delta^*$ state to find a decuplet~$\Xi^*$ with
similar properties.  

\subsection{\boldmath $\Xi$ spectrum and decays}

The $\Xi$~hyperons have the unique feature of double strangeness:
$|ssu\,\rangle$ and $|ssd\,\rangle$. If the confining potential is
independent of quark flavor, the energy of spatial excitations of a
given pair of quarks is inversely proportional to their reduced
mass. This means that the lightest excitations in each partial wave
are between the two strange quarks. In a spectator decay model, such
states will not decay to the ground state $\Xi$ and a pion because of
orthogonality of the spatial wave functions of the two strange quarks
in the excited state and the ground state. This removes the decay
channel with the largest phase space for the lightest states in each
partial wave, substantially reducing their widths. Typically,
$\Gamma_{\Xi^\ast}$ is about $10-20$~MeV for the known lower-mass
resonances, which is $5-30$~times narrower than for $N^\ast$,
$\Delta^\ast$, $\Lambda^\ast$, and $\Sigma^\ast$~states. These features, combined
with their isospin of 1/2, render possible a wide-ranging program on
the physics of the $\Xi$ hyperon and its excited states. Until recently, the study
of these hyperons has centered on their production in
$K^- p$~reactions; some $\Xi^\ast$~states were found using high-energy
hyperon beams. Photoproduction appears to be a very promising 
alternative. Results from earlier kaon beam experiments indicate that
it is possible to produce the $\Xi$~ground state through the decay of
high-mass hyperon and $\Xi$ states~\cite{Tripp:1967kj,Burgun:1969ee,Litchfield:1971ri}.
It is therefore possible to produce $\Xi$ resonances through 
$t$-channel photoproduction of hyperon resonances using the
photoproduction reaction $\gamma p\to K
K\,\Xi^{(\ast)}$~\cite{Price:2004xm,Guo:2007dw}. 

In briefly summarizing a physics program on Cascades, it would be
interesting to see the lightest excited $\Xi$~states in certain
partial waves decoupling from the $\Xi\pi$~channel, confirming the
flavor independence of confinement. Measurements of the isospin
splittings in spatially excited $\Xi$ states are also possible for
the first time. Currently, these
splittings, like $n-p$ or $\Delta^0 - \Delta^{++}$, are only available
for the octet and decuplet ground states, but are hard to measure in
excited $N,~\Delta$ and $\Sigma,~\Sigma^\ast$~states, which are very
broad. The lightest $\Xi$ baryons are expected to be narrower, and
measuring the $\Xi^- - \Xi^0$ splitting of spatially excited
$\Xi$~states remains a strong possibility. These measurements are an
interesting probe of excited hadron structure and would provide
important input for quark models, which describe the isospin splittings
by the $u$- and $d$-quark mass difference as well as by the
electromagnetic interactions between the quarks.

\subsection{\gx~sensitivity to $\Xi$ states}

The Cascades appear to be produced via $t$-channel exchanges that result
in the production of a hyperon $Y^{*}$ at the lower vertex which then decays to $\Xi^{(*)} K$.
Most of the momentum of the beam is transferred to a forward-going kaon that
is produced at the upper vertex.

The $\Xi$ octet ground states $(\Xi^0,\,\Xi^-)$ will be challenging to study in
the \gx~experiment via exclusive $t$-channel (meson exchange) production. The typical final
states for these studies,
\begin{eqnarray}
\label{Equation:GroundOctet}
\gamma p\,\to\, K\,Y^\ast\,&\to&\,K^+\, (\,\Xi^-\,K^+\,), \nonumber\\
 &~& K^+\,(\,\Xi^0\,K^0\,), \nonumber\\
 &~& K^0\,(\,\Xi^0\,K^+\,),
\end{eqnarray}
have kinematics for which the baseline \gx{} detector has very low acceptance due
to the high-momentum forward-going kaon and relatively low-momentum
pions produced in the $\Xi$ decay. 

However, the production of the $\Xi$ decuplet ground state, $\Xi(1530)$, and other
excited $\Xi$'s decaying to $\Xi\pi$ results in a lower momentum kaon at
the upper vertex, and these heavier $\Xi$ states produce higher momentum
pions in their decays.  \gx~will be able to search for and study excited $\Xi$'s 
in the reactions
\begin{eqnarray}
\label{Equation:GroundDecuplet}
\gamma p\,\to\, K\,Y^\ast\,&\to&\,K^+\,(\,\Xi\,\pi\,)\,K^0,\nonumber\\
&~&K^+\,(\,\Xi\,\pi\,)\,K^+,\nonumber\\ 
&~& K^0\,(\,\Xi\,\pi\,)\,K^+\,.
\end{eqnarray}

The lightest
excited $\Xi$~states are expected to decouple from $\Xi\pi$ and can
be searched for and studied also in their decays to $\Lambda \bar{K}$ and $\Sigma \bar{K}$:
\begin{eqnarray}
\label{Equation:Excited}
\gamma p\,\to\, K\,Y^\ast\,&\to&\,K^+\,(\,\bar{K}\Lambda\,)_{\Xi^{-\ast}}\,K^+,\nonumber\\
&~&K^+\,(\,\bar{K}\Lambda\,)_{\Xi^{0\ast}}\,K^0 ,\nonumber\\
&~&K^0\,(\,\bar{K}\Lambda\,)_{\Xi^{0\ast}}\,K^+,\\[1ex]
\gamma p\,\to\, K\,Y^\ast\,&\to&\,K^+\,(\,\bar{K}\Sigma\,)_{\Xi^{-\ast}}\,K^+,\nonumber\\
&~&K^+\,(\,\bar{K}\Sigma\,)_{\Xi^{0\ast}}\,K^0 ,\nonumber\\
&~&K^0\,(\,\bar{K}\Sigma\,)_{\Xi^{0\ast}}\,K^+.
\end{eqnarray}

\begin{figure}
\begin{center}
\includegraphics[width=.9\linewidth]{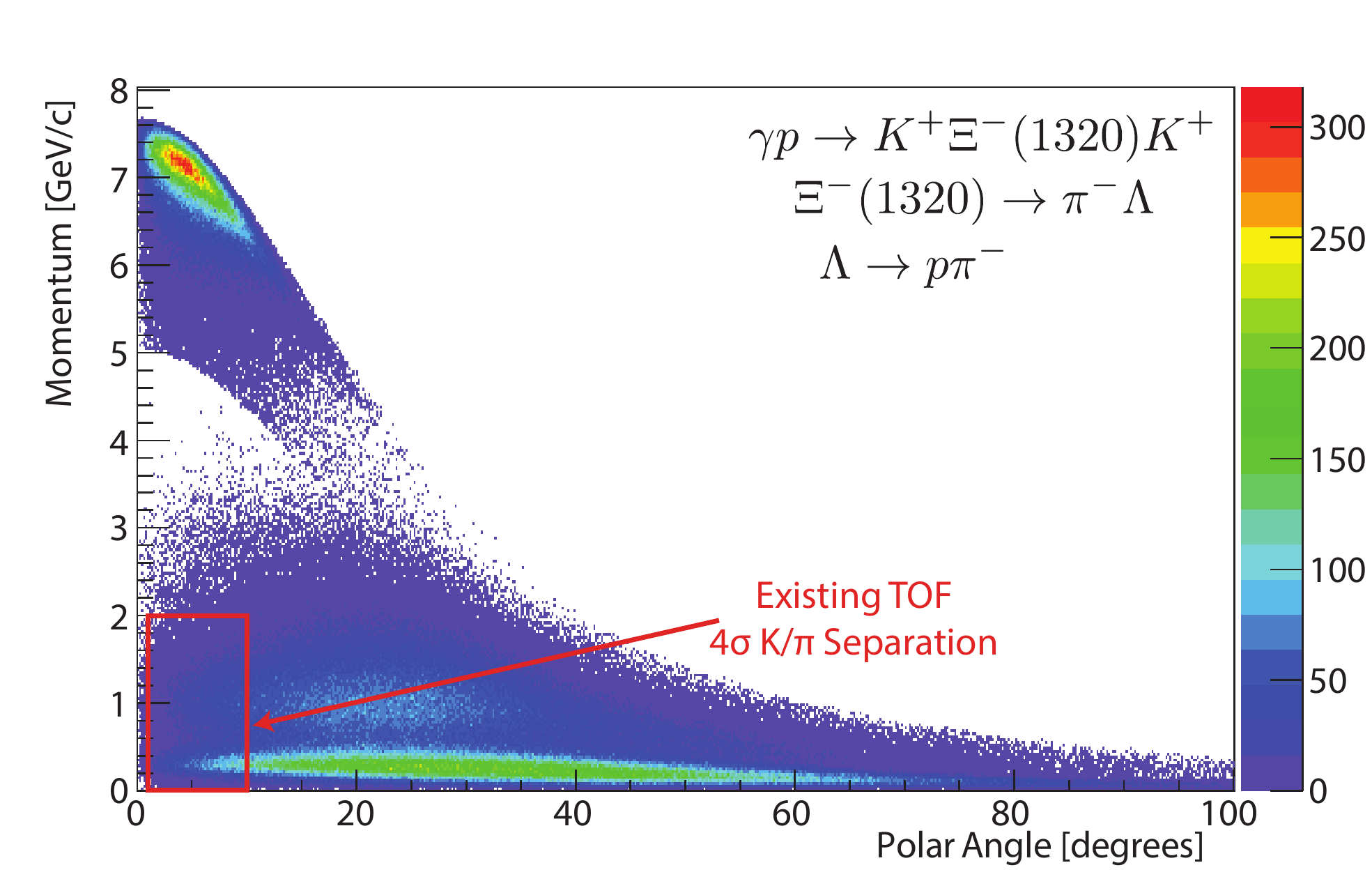}
\includegraphics[width=.9\linewidth]{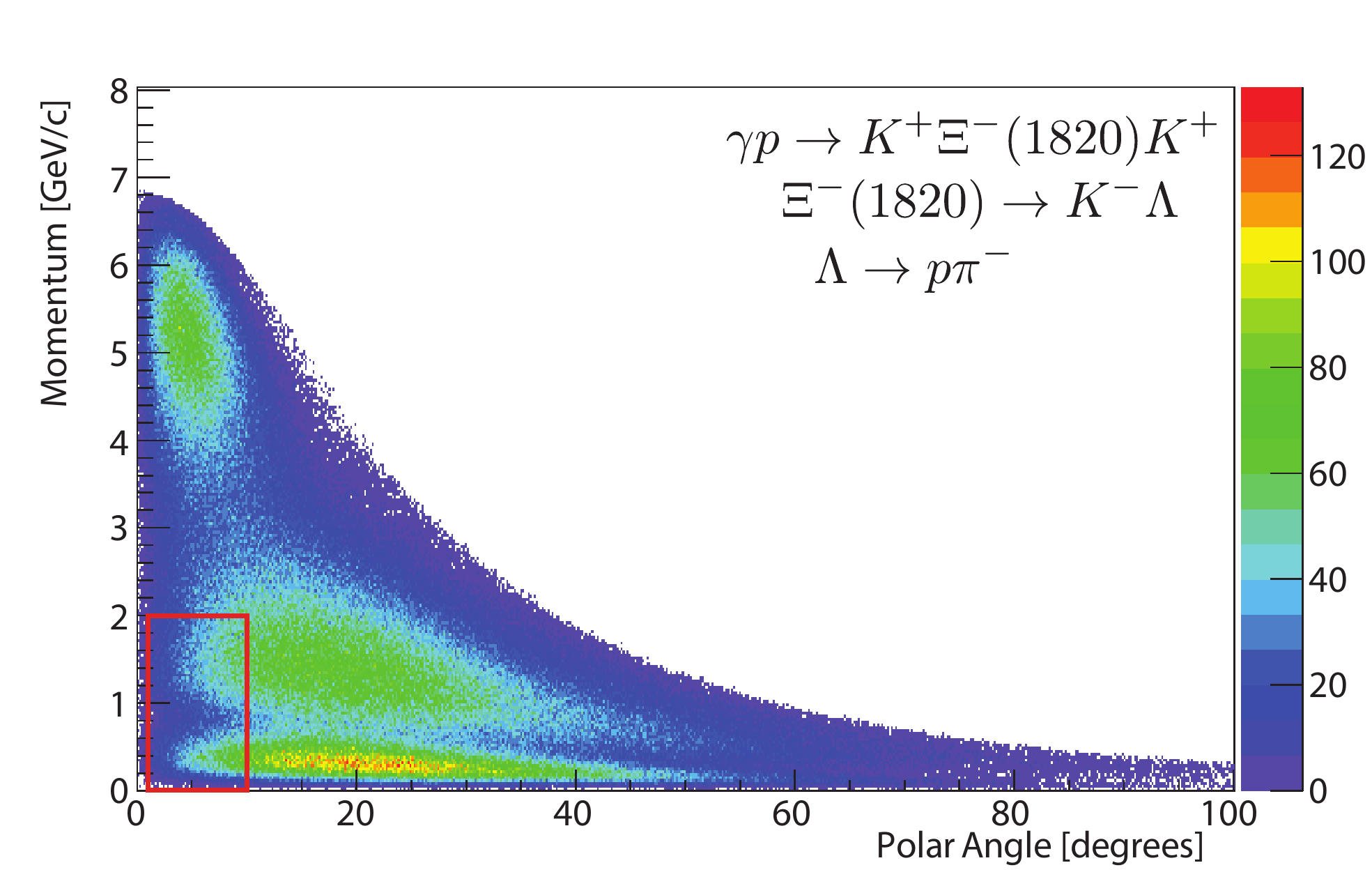}
\caption{\label{Figure:Production} Generated momentum versus polar
    angle for all tracks in the simulated reactions (top) $\gamma p\to
    K^+\,\Xi^-(1320) K^+$ and (bottom) $\gamma p\to K^+\,\Xi^-(1820)
    K^+$. The three high-density regions in each plot are populated with,
    from lowest to highest momentum, pions, kaons and protons, and kaons.}
\end{center}
\end{figure}

We have simulated the production of the $\Xi^-(1320)$ and
$\Xi^-(1820)$~resonances to better understand the kinematics of these
reactions. The photoproduction of the $\Xi^-(1320)$ decaying to
$\pi^-\Lambda$ and of the $\Xi^-(1820)$ decaying to $\Lambda K^-$ are
shown in Fig.~\ref{Figure:Production}. These reactions result in
$K^+K^+\pi^-\pi^-p$ and $K^+K^+K^-\pi^-p$ final states,
respectively. Reactions involving excited $\Xi$ states have lower-momentum
forward-going kaons,  making them more favorable for
study without supplemental particle ID hardware in the forward direction.
In addition, there is more energy available on average to
the $\Xi$ decay products, which results in a better detection efficiency
for the produced pions.

Using a BDT for signal selection, we have studied the specific reaction 
\begin{eqnarray*}
\gamma p & \to & K^+\Xi^{-}(1820) K^{+} \, ,
\end{eqnarray*}
with the subsequent decay of the $\Xi$ via 
\begin{eqnarray*}
\Xi^{-}(1820) & \to & \Lambda K^{-} \\
                      & \to & (p\pi^{-}) K^{-} \, .
\end{eqnarray*}
The signal selection efficiencies (post-reconstruction) are shown in Table~\ref{tab:bdt_eff_xi}.
As with the mesons, we find the efficiency should be adequate for conducting
a study of excited $\Xi$ states using the existing \gx~hardware.  Detailed studies
of the production, especially of the ground state $\Xi$'s, and a parity measurement
will likely require enhanced kaon identification in the forward direction, 
as presented in our PAC 39 proposal.

\begin{table}[h!]\centering
\caption[]{\label{tab:bdt_eff_xi}Selection efficiencies for identifying the $\Xi^{-}(1820)$. The efficiencies do 
not include the reconstruction efficiency of the final state tracks.}
\begin{tabular}{cccc} \hline\hline
~ & Tracking & Signal & Selection \\
Baryon & Resolution & Purity & Efficiency \\ \hline
$\Xi^{-}(1820)$   & Nominal & $0.90$ & $0.36$ \\
$\Xi^{-}(1820)$   & Nominal & $0.95$ & $0.27$ \\
\hline\hline
\end{tabular}
\end{table}

\section{\gx~Hardware and Beam Time Requirements}

In order to maximize the discovery capability of \gx, an increase in statistical 
precision beyond that expected from initial running is needed. In this section, we detail those 
needs. To maximize sensitivity, we propose a gradual increase in the photon flux towards the 
\gx~design of $10^8~\gamma/$s in the peak of the coherent bremsstrahlung spectrum 
($8.4~\mathrm{GeV} < E_\gamma < 9.0~\mathrm{GeV}$). Yield estimates, assuming an average 
flux of $5\times10^7~\gamma/$s, are presented.  In order to minimize the bandwidth
to disk and ultimately enhance analysis efficiency, we propose the addition of
a level-three software trigger to the \gx~data acquisition system.  The 
\gx~detector is designed to handle a rate of $10^8~\gamma/$s;
however, the optimum photon flux for taking data will depend on the beam condition and pileup at 
high luminosity and needs to be studied under realistic experimental conditions.  If our extraction of 
amplitudes is not limited by statistical uncertainties, we may optimize the flux to reduce systematic errors.

\subsection{\label{sec:level3}Level-three trigger}

The energy spectrum of photons striking the target ranges from near zero to the full $12$~GeV incident 
electron energy.  For physics analyses, one is primarily interested in only those events in the coherent 
peak around $9$~GeV, where there is a signal in the tagger that determines the photon energy.  At a 
rate of $10^7~\gamma$/s, the $120$~$\mu$b total hadronic cross section at $9$~GeV corresponds 
to a tagged hadronic event rate of about $1.5$~kHz.  Based on knowledge of the inclusive photoproduction 
cross section as a function of energy, calculations of the photon intensity in the region outside the
 tagger acceptance, and estimates for the  trigger efficiency, a total trigger rate of about $20$~kHz is 
expected.  At a typical raw event size of $15$~kB, the expected data rate of $300$~MB/s will saturate 
the available bandwidth to disk; rates higher than $10^7~\gamma$/s cannot be accommodated with 
the current data acquisition configuration.

For the high-intensity running, we propose the development of a level-three software trigger to loosely 
skim events that are consistent with a high energy $\gamma p$ collision.  The events of interest will be 
characterized by high-momentum tracks and large energy deposition in the calorimeter.  Matching observed
energy with a tagger hit is a task best suited for software algorithms like those used in physics analysis.  
It is expected that a processor farm can analyze multiple events in parallel, providing a real time
background rejection rate of at least a factor of ten.  While the exact network topology and choice of hardware will 
ultimately depend on the speed of the algorithm, at 10$^8~\gamma/$s the system will need to accommodate 
$3$~GB/s input from data crates, separate data blocks into complete events, and output the accepted events to disk at a rate 
of $<300$~MB/s.  The software trigger has the added advantage of increasing the concentration of 
tagged $\gamma p$ collision events in the output stream, which better optimizes use of disk
resources and enhances analysis efficiency.  Members of the \gx{} collaboration have developed and
implemented the software trigger for the LHCb experiment, which
is one of the most sophisticated software triggers ever developed~\cite{ref:bbdt,ref:roe}. 
We expect to benefit greatly from this expertise in developing an optimal level-three trigger for \gx.

The present baseline data acquisition system has been carefully developed so that a level-three software
trigger can be easily accommodated in the future.  We expect to begin prototyping the level-three
trigger using surplus computing hardware during the initial phases of \gx~running. This early
testing of both algorithms and hardware will allow us to specify our resource needs with good
accuracy in advance of the proposed Phase~IV running.

A simple estimate indicates that the implementation of a level-three trigger easily results in a net cost
savings rather than a burden.  Assuming no bandwidth limitations, if we write the entire unfiltered 
high-luminosity data stream to tape, the anticipated size is about $30$ petabytes per year
\footnote{This is at the \gx~design intensity of $10^8~\gamma/$s, which is
higher than our Phase IV average rate of $5\times10^7$ by a factor of two; however, other factors 
that would increase the data volume, such as event size increases due to higher-than-estimated 
noise, have not been included.}.  
Estimated media costs for storage of this data at the time of running would
be \$300K, assuming that no backup is made.  A data volume of this size would require the 
acquisition of one or more additional tape silos at a cost of about \$250K each.  Minimum storage costs
for a multi-year run will be nearing one million dollars.  Conversely, if we assume a level-three trigger 
algorithm can run a factor of ten faster than our current full offline reconstruction, then we can process 
events at a rate of 100~Hz per core.  The anticipated peak high luminosity event rate of 200~kHz would 
require 2000 cores, which at {\em today's costs} of 64-core machines would be about \$160K.  Even if 
a factor of two in computing is added to account for margin and data input/output overhead, the cost
is significantly lower than the storage cost. Furthermore, it is a fixed cost that does not grow with 
additional integrated luminosity, and it reduces the processing cost of the final stored data set when a 
physics analysis is performed on the data.

\subsection{Desired beam time and intensity}
\label{sec:beamtime}

There are several considerations in determining how much data one needs in any particular final state.
In order to perform an amplitude analysis of the final state particles (necessary for extracting
the quantum numbers of the produced resonances), one typically separates the data 
into bins of momentum transfer to the nucleon $t$ and resonance mass $M_X$.  The number of bins in $t$ could range 
from one to greater than ten, depending on the statistical precision of the data; a study of the 
$t$-dependence, if statistically permitted, provides valuable information on the production dynamics 
of particular resonances.  One desires to make the mass bins as small as possible in order to maximize 
sensitivity to states that have a small total decay width; however, it is not practical to use a bin size that 
is smaller than the resolution of $M_X$, which is on the order of $10$~MeV/$c^2$. In each 
bin of $t$ and $M_X$, one then needs enough events to perform an amplitude analysis, which is about 
$10^4$.  Therefore, our general goal is to reach a level of at least $10^4$ events per $10$~MeV/$c^2$
mass bin.  With more statistics, we can divide the data into bins of $t$ to study the production mechanism;
with fewer statistics, we may merge mass bins, which ultimately degrades the resolution with which
we can measure the masses and widths of the produced resonances.

In order to estimate the total event yield for various reactions of interest, we assume
200 PAC days of beam for the proposed Phase~IV running with 80\% of the delivered beam
usable for physics analysis.  The average Phase~IV intensity is assumed to be $5\times10^7~\gamma/$s
in the coherent bremsstrahlung peak.  This represents an integrated yield of events that
is approximately one order of magnitude larger than our approved Phase~II and~III running, 
which utilizes 90 PAC days of beam for physics analysis\footnote{We plan to utilize 30 of the 120 
approved PAC days for the Phase~I commissioning of the detector.} at an average intensity
of $10^7~\gamma/$s in the coherent peak.  Table~\ref{tab:params} summarizes the various
running configurations.  

Below we present two independent estimates of event yields to justify our request
for 200 PAC days of beam.  Both reach similar conclusions:  the proposed run would provide
sufficient statistics to conduct an initial amplitude analysis of the mass spectrum for several
select $s\bar{s}$ meson decay modes.  In addition, the resulting order-of-magnitude increase
in statistical precision will allow a more detailed exploration of those topologies such as $\eta^\prime \pi$,
$b_1\pi$, or $f_1\pi$ that may be statistically limited in the initial \gx~running.  Finally, 
the spectrum of $\Xi$ baryons can also be studied with high statistical precision.

\begin{table*}
\begin{center}
\caption{\label{tab:params}A table of relevant parameters for the various phases of \gx~running.}
\begin{tabular}{l|ccc|c}\hline\hline
 & & Approved & & {\em Proposed} \\
 & Phase I & Phase II & Phase III & ~~Phase IV~~ \\ \hline 
 Duration (PAC days) & 30 & 30 & 60 & 200 \\
 Minimum electron energy (GeV) & 10 & 11 & 12 & 12 \\
 Average photon flux  ($\gamma$/s) & $10^6$ & $10^7$ & $10^7$ & $5 \times 10^7$ \\
 Average beam current (nA) & 50 - 200\footnote{An amorphous radiator may be used for some commissioning and later
 replaced with a diamond.}  & 220 & 220 & 1100 \\
 Maximum beam emittance (mm$\cdot\mu$r) & 50 & 20 & 10 & 10 \\
 Level-one (hardware) trigger rate (kHz) & 2 & 20 & 20 & 200 \\
 Raw Data Volume (TB) & 60 & 600 & 1200 & 2300\footnote{This volume assumes the implementation of the 
 proposed level-three software trigger.} \\ \hline\hline
 \end{tabular}
 \end{center}
 \end{table*}

\subsubsection{Meson yields based on cross section estimates}

One can estimate the total number of observed events, $N_{i}$, in some stable final state 
by
\begin{equation}
N_i = \epsilon_i \sigma_i n_\gamma n_t T,
\label{eq:rate}
\end{equation}
where $\epsilon_i$ and $\sigma_i$ are the detection efficiency and 
photoproduction cross section of the final state $i$, $n_\gamma$ is the rate 
of incident photons on target, $n_t$ is the number of scattering centers per 
unit area, and $T$ is the integrated live time of the detector.  For a 30~cm 
LH$_2$ target, $n_t$ is 1.26~b$^{-1}$.  (A useful rule of thumb is that at 
$n_\gamma = 10^7~\gamma$/s a 1~$\mu$b cross section will result in the 
production of about $10^{6}$ events per day.) 
It is difficult to estimate the production cross section for many final states 
since data in the \gx~energy regime are sparse.  (For a compendium of photoproduction 
cross sections, see Ref.~\cite{Baldini:1988ti}.)  Table~\ref{tab:yields} lists key final states
for initial exotic hybrid searches along with assumed production cross sections\footnote{Some 
estimates are based on actual data from Ref.~\cite{Baldini:1988ti} for cross sections at a similar 
beam energy, while others are crudely estimated from the product of branching ratios of heavy 
meson decays, {\it i.e.,} a proxy for light meson hadronization ratios, and known photoproduction 
cross sections.}.

\begin{figure}
 \begin{center}
 \includegraphics[width=\linewidth]{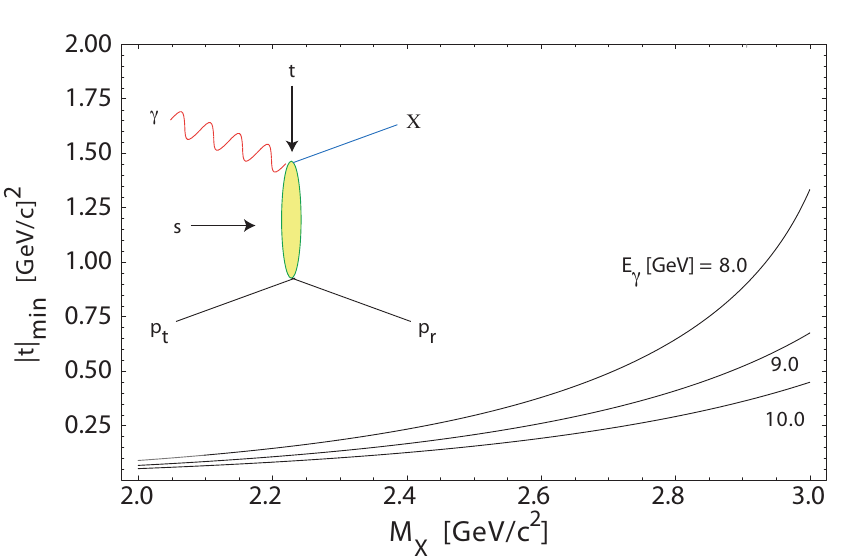}
 \caption{\label{fig:tmin}Dependence of $|t|_\mathrm{min}$ on the mass of the outgoing meson system $M_X$.  The lines indicate incident photon energies of 8.0, 9.0, and 10.0 GeV.}
\end{center}
 \end{figure}

\begin{figure}
 \begin{center}
 \includegraphics[width=0.85\linewidth]{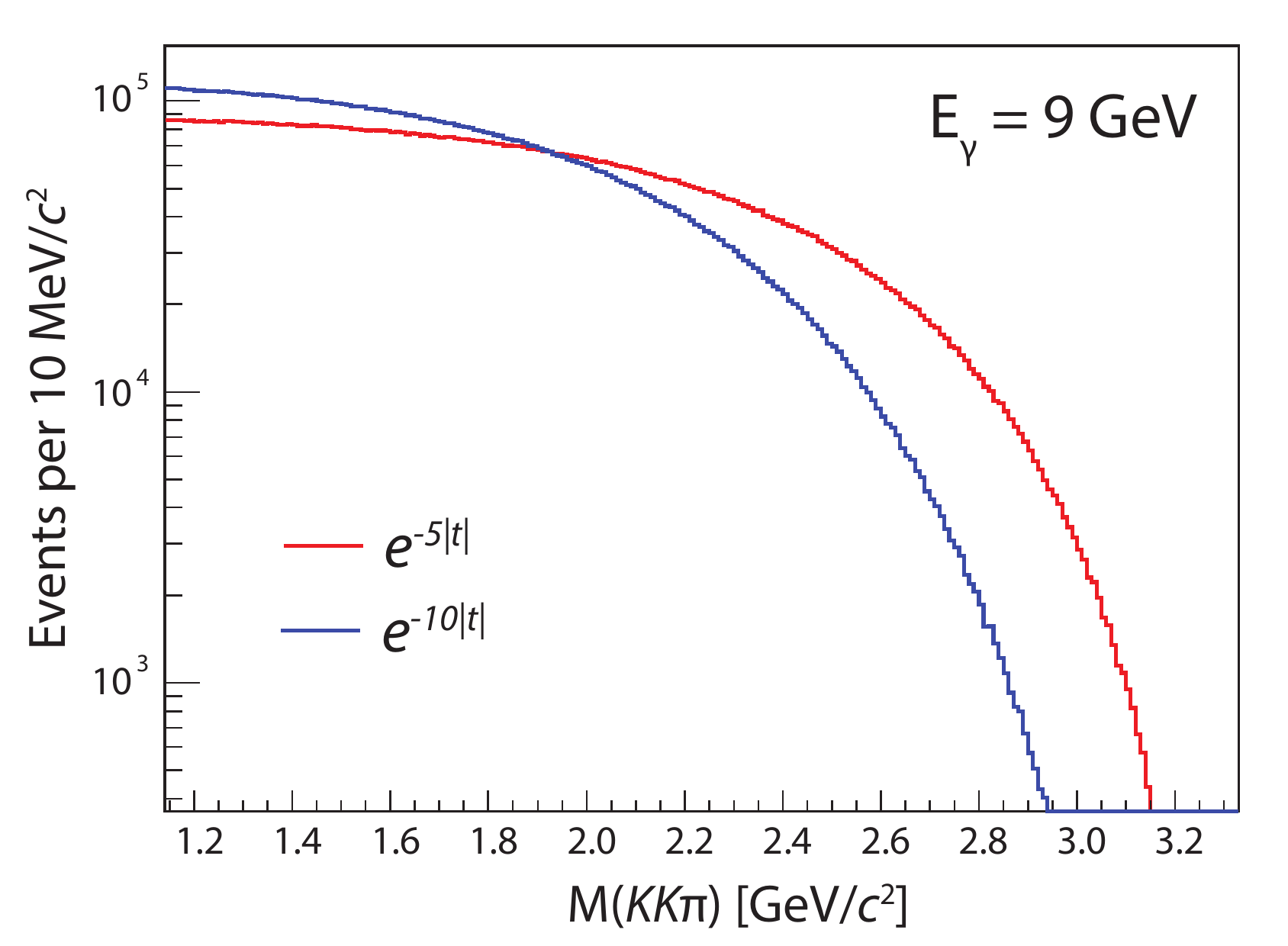}
 \caption{\label{fig:n_vs_mass}A figure showing the number of expected events per 10~MeV/$c^2$ bin in the $KK\pi$ invariant mass distribution, integrating over all allowed values of $t$, and assuming $10^7$ events in total are detected.  No dependence on $M(KK\pi)$ is assumed, although, in reality, the mass dependence will likely be driven by resonances.  Two different assumptions for the $t$ dependence are shown.  The region above 2.5~GeV/$c^2$ represents about 8\% (2\%) of all events for the $\alpha = 5 (10)~($GeV$/c)^{-2}$ values.}
\end{center}
 \end{figure}

Photoproduction of mesons at $9$~GeV proceeds via peripheral production (sketched 
in the inset of Fig.~\ref{fig:tmin}). The production can typically be characterized as a function of 
$t\equiv (p_X-p_\gamma)^2$, with the production cross section proportional to $e^{-\alpha|t|}$.  The 
value of $\alpha$ for measured reactions ranges from 3 to 10~GeV$^{-2}$. This $t$-dependence, 
which is unknown for many key \gx~reactions, results in a suppression of the rate at large values 
of $|t|$, which, in turn, suppresses the production of high mass mesons.  Figure~\ref{fig:tmin} shows 
the minimum value of $|t|$ as a function of the produced meson mass $M_X$ for a variety of different 
photon energies.  The impact of this kinematic suppression on a search for heavy states is illustrated 
in Figure~\ref{fig:n_vs_mass},  where events are generated according to the $t$ distributions with 
both $\alpha=5~($GeV$/c)^{-2}$ and $10~($GeV$/c)^{-2}$ and uniform in $M_X$.  Those that are kinematically 
 allowed ($|t|>|t|_\mathrm{min}$) are retained.  The $y$-axis indicates the number of events in 
$10$~MeV/$c^2$ mass bins, integrated over the allowed region in $t$, and assuming a total of 
$3\times 10^7$ events are collected.  The region above $M_X=2.5~$GeV$/c^2$, where one would 
want to search for states such as the $h_{2}$ and $h_{2}^{\prime}$, contains only about 5\% of all events 
due to the suppression of large $|t|$ that is characteristic of peripheral photoproduction.
  
To estimate our total yield in various final states, we assume the detection efficiency for 
protons, pions, kaons, and photons to be 70\%, 80\%, 40\%, and 
80\%, respectively.  Of course, the true efficiencies are dependent on software algorithms, kinematics, 
multiplicity, and other variables; however, the dominant uncertainty in yield estimates is not efficiency 
but cross section.  These assumed efficiencies roughly reproduce signal selection efficiencies in detailed 
simulations of $\gamma p \to \pi^+\pi^-\pi^+ n$, $\gamma p \to \eta \pi^0 p$, 
$\gamma p \to b_1^\pm \pi^\mp p$, and $\gamma p \to f_1\pi^0 p$ performed by the collaboration, as
well as the BDT selection efficiencies presented earlier.  Table~\ref{tab:yields} provides an estimate
of the detected yields for various topologies for our proposed Phase~IV run.
If we take the $KK\pi$ channel as an example, Figure~\ref{fig:n_vs_mass} demonstrates that, under 
some assumptions about the production, the proposed run yields enough statistics to
just meet our goal of $10^4$ events per mass bin in the region where $s\bar{s}$ exotics are 
expected to reside.  

\begin{table}
\begin{center}
\caption{\label{tab:yields}A table of hybrid search channels, estimated cross sections, and approximate numbers of observed events for the proposed Phase~IV running.  See text for a discussion of the underlying assumptions.  The subscripts on $\omega$, $\eta$, and $\eta^\prime$ indicate the decay modes used in the efficiency calculations.  If explicit charges are not indicated, the yields represent an average over various charge combinations.}
\begin{tabular}{cccc}\hline\hline 
& Cross & Proposed \\
Final &  ~~Section~~ & Phase IV \\
  State & ($\mu$b) & ($\times10^6$ events) \\  \hline
$\pi^+\pi^-\pi^+$ & 10 & 3000 \\
$\pi^+\pi^-\pi^0$ & 2 & 600 \\
$KK\pi\pi$ & 0.5 & 40 \\
$KK\pi$ & 0.1 & 10 \\ 
$\omega_{3\pi}\pi\pi$ & 0.2 & 40 \\
$\omega_{\gamma\pi}\pi\pi$ & 0.2 & 6 \\
$\eta_{\gamma\gamma}\pi\pi$ & 0.2 & 30 \\
$\eta_{\gamma\gamma}\pi\pi\pi$ & 0.2 & 20 \\
$\eta^\prime_{\gamma\gamma}\pi$ & 0.1 & 1 \\
$\eta^\prime_{\eta\pi\pi}\pi$ & 0.1 & 3 \\ \hline \hline
\end{tabular}
\end{center}
\end{table}

\subsubsection{Meson yields based on \textsc{pythia} simulation}

\begin{table*}\centering
\caption[]{\label{tab:pythia_rates}\textsc{pythia}-predicted numbers of events for various exclusive final states in
a mass range appropriate for searching for various mesons.  Estimates are based on $200$ PAC
days at $80$\% uptime at an average intensity of $5\times 10^{7}\gamma /$s.  Events per
10~MeV/$c^2$ is an estimate of the number of events available for an amplitude
analysis in each mass bin.} 
\begin{tabular}{llcccc} \hline\hline
~~~Meson of &  Reaction & \multicolumn{2}{c}{Mass Range [MeV/$c^2$]} & Signal & Events per \\
~~~Interest ($X$) ~~~~& Topology &~~$M_X^\mathrm{min}$ & ~~$M_X^\mathrm{max}$ &~~~Yield [10$^6$] ~~~&~~10 MeV/$c^2$ [10$^4$] \\ \hline
$h_2^\prime(2600)$ & $\gamma p \to (K_1(1400) K)_X p~~~~$ & 2415 & 2785 &  1.5 &  4.0 \\
 ~ & ~~$K_1\to K^*\pi$ \\
 ~ & ~~~~$K^* \to K\pi$ \\
$\eta_1^\prime(2300)$ & $\gamma p \to (K^* K_S )_X p$ & 2000 & 2600 & 0.46 & 1.5 \\
~ & ~~$K^* \to K^\pm \pi^\mp$ \\
~ & ~~$K_S\to \pi^+ \pi^-$ \\
$\phi_3(1850) $ & $\gamma p \to (K^+ K^-)_X p$ & 1720 & 1980 & 5.3 & 21\\
$Y(2175)$ & $\gamma p \to (\phi f_0(980) )_X p$ & 2060 & 2290 & 0.12 & 0.52 \\ 
~ & ~~$\phi\to K^+K^-$ \\
~ & ~~$f_0(980)\to K^+K^-$ \\
\hline \hline
\end{tabular}
\end{table*}

We have also used \textsc{pythia} to simulate the expected yields of various hadronic final 
states.  \textsc{pythia} reproduces known photoproduction cross sections relatively well; therefore,
it is expected to be an acceptable estimator of the production rates of various topologies
where we would like to search for new mesons.  Using the large $5\times 10^9$ event
inclusive-photoproduction \textsc{pythia} sample, we can analyze the signal yield when we attempt
to reconstruct and select various final state topologies.  The signal selection is
performed using a BDT, as discussed earlier, with a goal of 90\% signal purity.  We place
loose requirements on the invariant masses of the intermediate resonances.  The
measured yield after reconstruction and selection can then be scaled to estimate
the number of reconstructed signal events that our Phase~IV running would produce.
In Table~\ref{tab:pythia_rates} we show the various topologies studied.  In addition,
we measure the yield in a region of meson candidate $X$ invariant mass to estimate
the statistical precision of an amplitude analysis in that region.  The number of events
per 10~MeV/$c^2$ mass bin is listed, and we observe that we meet our goal
of $10^4$ events per bin in most topologies.  The $K^*K_S$ yield in Table~\ref{tab:pythia_rates}
also loosely agrees with that in Figure~\ref{fig:n_vs_mass}, and the ratio of $K_1K$ to $K^*K$
roughly matches that obtained by estimated cross sections and efficiencies as shown in
Table~\ref{tab:yields}.

\subsubsection{$\Xi$ yields}

Existing knowledge of $\Xi$ photoproduction can be used to estimate the expected
yields of $\Xi$ states.
Recent CLAS data for the $\Xi(1320)$ are consistent with $t$-slope values ranging from 
1.11 to 2.64~$({\rm GeV}/c)^{-2}$ for photon energies between 2.75 and 3.85~GeV~\cite{Guo:2007dw}. 
Values for excited $\Xi$'s are not well known, but a recent unpublished CLAS analysis 
of a high-statistics data sample (JLab proposal E05-017) indicates that the $t$-slope value 
flattens out above 4~GeV at a value of about 1.7~$({\rm GeV}/c)^{-2}$~\cite{GoetzThesis}. 
We have used a value of 1.4~$({\rm GeV}/c)^{-2}$ for the $\Xi^-(1320)$ and 1.7~$({\rm GeV}/c)^{-2}$ 
for the $\Xi^-(1820)$ in our simulations at 9 GeV. The most recent published CLAS analysis~\cite{Guo:2007dw} 
has determined a total cross section of about 15~nb and 2~nb for the $\Xi^-(1320)$ 
and $\Xi^-(1530)$, respectively, at $E_\gamma=5$~GeV. An unpublished analysis shows that the 
total cross section levels out above 3.5~GeV, but the energy range is limited at 
5.4~GeV~\cite{GoetzThesis}. A total number of about 20,000 $\Xi^-(1320)$~events 
was observed for the energy range $E_\gamma\in [2.69,\,5.44]$~GeV. 

The BDT analysis carried out using $K^+K^+K^-\Lambda$ \textsc{pythia} signal events suggests that the 
proposed \gx{} run will result in a yield of about 90,000 of these events with 90\% purity for 
$K^-\Lambda$ invariant mass in the $\Xi$(1820) mass region.  Estimates using Eq.~(\ref{eq:rate}) lead us to 
expect yields of about $800,000$ $\Xi^-(1320)$ and $100,000$ $\Xi^-(1530)$ events.
Such high statistics samples of exclusively reconstructed $\Xi$ final states greatly
enhance the possibility of determining the spin and parity of excited states.

In summary, we request a production run consisting of $200$ days of beam time at 
an average intensity of $5\times 10^7~\gamma$/s for production Phase~IV running of the 
\gx~experiment. It is anticipated that the Phase~IV intensity will start around $10^7~\gamma/$s, our 
Phase~III intensity, and increase toward the \gx~design goal of $10^8~\gamma/$s as we understand 
the detector capability for these high rates based on the data acquired at $10^7~\gamma/$s.  The data 
sample acquired will provide an order of magnitude improvement in statistical precision over the initial 
Phase II and III running of \gx, which will allow an initial study of high-mass states containing strange
quarks and an exploration of the $\Xi$ spectrum.

\section{Summary}

We propose an expansion of the present baseline \gx~experimental program to include
an order-of-magnitude higher statistics by increasing the average tagged photon intensity by
a factor of five and the beam time by a factor of two.  The increase in intensity necessitates
the implementation of the \gx~level-three software trigger.  The program requires 200 days of
beam time with 9~GeV tagged photons at an average intensity of $5\times 10^7~\gamma/s$. 
We have demonstrated that the baseline \gx~detector design is capable of reconstructing
particular final state topologies that include kaons.  While the acceptance and purity may be
limited without the addition of supplemental kaon identification hardware, the proposed
run will provide a level of statistical precision sufficient to make an initial study of meson 
states with an $s\bar{s}$ component and to search for excited doubly-strange $\Xi$-baryon
states.


\begin{thebibliography}{2}

\bibitem{Dudek:2011bn} 
  J.~J.~Dudek,
  Phys.\ Rev.\ D {\bf 84}, 074023 (2011).
  
 \bibitem{pac30}
  \gx~Collaboration, ``Mapping the Spectrum of Light Quark Mesons and Gluonic Excitations with Linearly Polarized Protons," {\it Presentation to PAC 30}, (2006).  Available at:  \href{http://www.gluex.org/docs/pac30_proposal.pdf}{\it http://www.gluex.org/docs/pac30\_proposal.pdf}
  
  \bibitem{pac36}
  \gx~Collaboration,  ``The \gx~Experiment in Hall D," {\it Presentation to PAC 36}, (2010).  Available at: \href{http://www.gluex.org/docs/pac36_update.pdf}{\it http://www.gluex.org/docs/pac36\_update.pdf}

  \bibitem{pac39}
  \gx~Collaboration,  ``A study of meson and baryon decays to strange final states with \gx~in Hall D," {\it Presentation to PAC 39}, (2012).  Available at: \href{http://www.gluex.org/docs/pac39_proposal.pdf}{\it http://www.gluex.org/docs/pac39\_proposal.pdf}
  
  \bibitem{Dudek:2009qf} 
  J.~J.~Dudek, R.~G.~Edwards, M.~J.~Peardon, D.~G.~Richards and C.~E.~Thomas,
  Phys.\ Rev.\ Lett.\  {\bf 103}, 262001 (2009).
  
  \bibitem{Dudek:2010wm} 
  J.~J.~Dudek, R.~G.~Edwards, M.~J.~Peardon, D.~G.~Richards and C.~E.~Thomas,
  Phys.\ Rev.\ D {\bf 82}, 034508 (2010).
  
  \bibitem{Dudek:2011tt} 
  J.~J.~Dudek, R.~G.~Edwards, B.~Joo, M.~J.~Peardon, D.~G.~Richards and C.~E.~Thomas,
  Phys.\ Rev.\ D {\bf 83}, 111502 (2011).
  
  \bibitem{Edwards:2011jj} 
  R.~G.~Edwards, J.~J.~Dudek, D.~G.~Richards and S.~J.~Wallace,
  Phys.\ Rev.\ D {\bf 84}, 074508 (2011).
  
  \bibitem{Dudek:2012ag} 
  J.~J.~Dudek and R.~G.~Edwards,
  Phys.\ Rev.\ D {\bf 85}, 054016 (2012).

\bibitem{Edwards:2012fx} 
  R.~G.~Edwards, N.~Mathur, D.~G.~Richards and S.~J.~Wallace,
  Phys.\  Rev.\ D {\bf 87}, 054506 (2013)
  [arXiv:1212.5236 [hep-ph]].

\bibitem{Beringer:1900zz} 
  J.~Beringer {\it et al.}  [Particle Data Group Collaboration],
  Phys.\ Rev.\ D {\bf 86}, 010001 (2012).
  
  \bibitem{Ablikim:2010au} 
  M.~Ablikim {\it et al.}  [BESIII Collaboration],
  Phys.\ Rev.\ Lett.\  {\bf 106}, 072002 (2011).
  
  \bibitem{Adams:2011sq} 
  G.~S.~Adams {\it et al.}  [CLEO Collaboration],
  Phys.\ Rev.\ D {\bf 84}, 112009 (2011).
  
  \bibitem{Ivanov:2001rv} 
  E.~I.~Ivanov {\it et al.}  [E852 Collaboration],
  Phys.\ Rev.\ Lett.\  {\bf 86}, 3977 (2001).
  
  \bibitem{Meyer:2010ku} 
  C.~A.~Meyer and Y.~Van Haarlem,
  Phys.\ Rev.\ C {\bf 82}, 025208 (2010).
  
  \bibitem{Aubert:2005rm} 
  B.~Aubert {\it et al.}  [BABAR Collaboration],
  Phys.\ Rev.\ Lett.\  {\bf 95}, 142001 (2005).
  
  \bibitem{Coan:2006rv} 
  T.~E.~Coan {\it et al.}  [CLEO Collaboration],
  Phys.\ Rev.\ Lett.\  {\bf 96}, 162003 (2006).
  
  \bibitem{He:2006kg} 
  Q.~He {\it et al.}  [CLEO Collaboration],
  Phys.\ Rev.\ D {\bf 74}, 091104 (2006).
  
  \bibitem{Yuan:2007sj} 
  C.~Z.~Yuan {\it et al.}  [Belle Collaboration],
  Phys.\ Rev.\ Lett.\  {\bf 99}, 182004 (2007).
  
  \bibitem{Aubert:2006bu} 
  B.~Aubert {\it et al.}  [BABAR Collaboration],
  Phys.\ Rev.\ D {\bf 74}, 091103 (2006).
  
  \bibitem{Ablikim:2007ab} 
  M.~Ablikim {\it et al.}  [BES Collaboration],
  Phys.\ Rev.\ Lett.\  {\bf 100}, 102003 (2008).
  
  \bibitem{Shen:2009zze} 
  C.~P.~Shen {\it et al.}  [Belle Collaboration],
  Phys.\ Rev.\ D {\bf 80}, 031101 (2009).
  
\bibitem{Price:2004xm}
  J.~W.~Price {\it et al.}  [CLAS Collaboration],
  Phys.\ Rev.\  C {\bf 71}, 058201 (2005).

\bibitem{Guo:2007dw}
  L.~Guo {\it et al.},
  Phys.\ Rev.\  C {\bf 76}, 025208 (2007).

    \bibitem{Dudek:2008sz}  
  J.~J.~Dudek and E.~Rrapaj,
  Phys.\ Rev.\ D {\bf 78}, 094504 (2008).
  
  \bibitem{Liu:2012ze} 
  L.~Liu {\it et al.},
  arXiv:1204.5425 [hep-ph].

\bibitem{VanHaarlem:2010yq} 
  Y.~Van Haarlem {\it et al.},
  Nucl.\ Instrum.\ Meth.\ A {\bf 622}, 142 (2010).
  
  \bibitem{Brunner:1998fh} 
  A.~Brunner {\it et al.},
  Nucl.\ Instrum.\ Meth.\ A {\bf 414}, 466 (1998).
  
 \bibitem{FCALBeamTestNIM}
 K.~Moriya {\it et al.}, arXiv:1304.499[physics-ins.det] (submitted to Nucl.\ Instrum.\ Meth.\ A)
(2013).
  
\bibitem{ReviewReport}
David Nathan Brown, Sergei Gerassimov, Chris Jones, Martin Purschke, and Torre Wenaus,
``Report of the 12 GeV Software and Computing Review,''
GlueX-doc-2055, September 2012.

  
  \bibitem{Adams:1998ff} 
  G.~S.~Adams {\it et al.}  [E852 Collaboration],
  Phys.\ Rev.\ Lett.\  {\bf 81}, 5760 (1998).
  
  \bibitem{Dzierba:2005jg} 
  A.~R.~Dzierba {\it et al.},
  Phys.\ Rev.\ D {\bf 73}, 072001 (2006).
  
  \bibitem{Alekseev:2009aa} 
  M.~Alekseev {\it et al.}  [COMPASS Collaboration],
  Phys.\ Rev.\ Lett.\  {\bf 104}, 241803 (2010).
  
  \bibitem{Nozar:2008aa} 
  M.~Nozar {\it et al.}  [CLAS Collaboration],
  Phys.\ Rev.\ Lett.\  {\bf 102}, 102002 (2009).
  
  \bibitem{IgorThesis}
  I.~Senderovich, Ph.D. Thesis (unpublished), University of Connecticut (2012).
  
\bibitem{ref:bdt} L. Brieman {\em el al.}, {\em Classification and regression trees}, Wadsworth International Group, Belmont, California (1984).

\bibitem{ref:lhcbhlt} R. Aaij {\em et al.} [LHCb Trigger Group], {\em The LHCb trigger and its performance}, Submitted to JINST. [arxiv:1211.3055]

\bibitem{ref:bbdt}  V. Gligorov and M. Williams, {\em Efficient, reliable and fast high-level triggering using a bonsai boosted decision tree}, JINST {\bf 8}, P02013 (2013).

\bibitem{ref:roe} B.P. Roe {\em et al.}, {\em Boosted decision trees as an alternative to artificial neural networks for particle identification}, Nucl.Instrum.Meth.\ A {\bf 543}, 577 (2005).
  
  \bibitem{Close:2005iz} 
  F.~E.~Close and P.~R.~Page,
  Phys.\ Lett.\ B {\bf 628}, 215 (2005).

  \bibitem{Zhu:2005hp} 
  S.~-L.~Zhu,
  Phys.\ Lett.\ B {\bf 625}, 212 (2005).
  
  \bibitem{Kou:2005gt} 
  E.~Kou and O.~Pene,
  Phys.\ Lett.\ B {\bf 631}, 164 (2005).
    
    \bibitem{Luo:2005zg} 
  X.~-Q.~Luo and Y.~Liu,
  Phys.\ Rev.\ D {\bf 74}, 034502 (2006)
  [Erratum-ibid.\ D {\bf 74}, 039902 (2006)].
      
  \bibitem{Ding:2007pc} 
  G.~-J.~Ding and M.~-L.~Yan,
  Phys.\ Lett.\ B {\bf 657}, 49 (2007).
  
\bibitem{Page:1998gz} 
  P.~R.~Page, E.~S.~Swanson and A.~P.~Szczepaniak,
  Phys.\ Rev.\ D {\bf 59}, 034016 (1999).

\bibitem{Isgur:1985vy} 
  N.~Isgur, R.~Kokoski and J.~E.~Paton,
  Phys.\ Rev.\ Lett.\  {\bf 54}, 869 (1985).
  
    
\bibitem{ref:TMVA2007} A.~Hoecker, P.~Speckmayer, J.~Stelzer, J.~Therhaag, E.~von Toerne, and H.~Voss, {\em TMVA: Toolkit for Multivariate Data Analysis}, PoS A CAT 040 (2007). 

\bibitem{Aubert:2008ty} 
  B.~Aubert {\it et al.}  [BABAR Collaboration],
  Phys.\ Rev.\ D {\bf 78}, 034008 (2008).

\bibitem{Biagi:1986vs} 
  S.~F.~Biagi {\it et al.},
  Z.\ Phys.\ C {\bf 34}, 175 (1987).
\bibitem{Tripp:1967kj}
  R.~D.~Tripp {\it et al.}.,
  Nucl.\ Phys.\  B {\bf 3}, 10 (1967).

\bibitem{Burgun:1969ee} 
  G.~Burgun {\it et al.},
  Nucl.\ Phys.\  B {\bf 8}, 447 (1968).

\bibitem{Litchfield:1971ri}
  P.~J.~Litchfield {\it et al.},
  Nucl.\ Phys.\  B {\bf 30}, 125 (1971).

  \bibitem{Baldini:1988ti} 
  A.~Baldini, V.~Flaminio, W.~G.~Moorhead, D.~R.~O.~Morrison and H.~Schopper, (Ed.),
  ``Numerical Data And Functional Relationships In Science And Technology. Grp. 1: Nuclear And Particle Physics. Vol. 12: Total Cross-sections For Reactions Of High-energy Particles (including Elastic, Topological, Inclusive And Exclusive Reactions). Subvol,''
  BERLIN, GERMANY: SPRINGER (1988) 409 P. (LANDOLT-BOERNSTEIN. NEW SERIES, 1/12B)

\bibitem{GoetzThesis} John Theodore Goetz, Ph.D. Thesis, University of California, Los Angeles, 2010 (unpublished).

\end{thebibliography}
\end{document}